\documentclass[aps,prc,twocolumn,showpacs,superscriptaddress,fixfloats]{revtex4-1}
\usepackage[english]{babel}
\usepackage[T1]{fontenc}
\usepackage[colorlinks=true,linkcolor=blue,citecolor=blue,urlcolor=blue]{hyperref}
\usepackage{bm,bbm,amssymb,amsmath}
\usepackage{graphicx}
\usepackage{multirow}
\usepackage[capitalize]{cleveref}

\begin{document}

\title{Variational calculation of the ground state of closed-shell nuclei up to $A=40$}

\author{D. Lonardoni}
\email{lonardoni@nscl.msu.edu}
\affiliation{National Superconducting Cyclotron Laboratory, Michigan State University, East Lansing, Michigan 48824}
\affiliation{Theoretical Division, Los Alamos National Laboratory, Los Alamos, New Mexico 87545}

\author{A. Lovato}
\affiliation{Physics Division, Argonne National Laboratory, Argonne, Illinois 60439}

\author{Steven C. Pieper}
\affiliation{Physics Division, Argonne National Laboratory, Argonne, Illinois 60439}

\author{R. B. Wiringa}
\affiliation{Physics Division, Argonne National Laboratory, Argonne, Illinois 60439}

\begin{abstract}
Variational calculations of ground-state properties of $^4$He, $^{16}$O, and $^{40}$Ca are carried out employing realistic phenomenological two- and three-nucleon potentials. The trial wave function includes two- and three-body correlations acting on a product of single-particle determinants. Expectation values are evaluated with a cluster expansion for the spin-isospin dependent correlations considering up to five-body cluster terms. The optimal wave function is obtained by minimizing the energy expectation value over a set of up to 20 parameters by means of a nonlinear optimization library. We present results for the binding energy, charge radius, one- and two-body densities, single-nucleon momentum distribution, charge form factor, and Coulomb sum rule. We find that the employed three-nucleon interaction becomes repulsive for $A\geq16$. In $^{16}$O the inclusion of such a force provides a better description of the properties of the nucleus. In $^{40}$Ca instead, the repulsive behavior of the three-body interaction fails to reproduce experimental data for the charge radius and the charge form factor. We find that the high-momentum region of the momentum distributions, determined by the short-range terms of nuclear correlations, exhibit a universal behavior independent of the particular nucleus. The comparison of the Coulomb sum rules for $^4$He, $^{16}$O, and $^{40}$Ca reported in this work will help elucidate in-medium modifications of the nucleon form factors.
\end{abstract}

\maketitle

\section{Introduction}
\label{sec:intro}
Atomic nuclei are self-bound systems of strongly interacting fermions. Understanding their structure, reactions, and electroweak properties in terms of the individual interactions among their constituents, protons and neutrons, has been a long-standing goal of theoretical nuclear physics. Ab initio approaches are aimed at solving the many-body Schr\"odinger equation associated with the nuclear Hamiltonian. This is made particularly difficult by the strong coupling of spin and spatial degrees of freedom which characterize nuclear forces. In addition, the nuclear many-body solution has to feature a self-emerging shell structure and should be able to encompass clusters of highly correlated nucleons.

One of the key advantages of ab initio approaches is that they allow the disentanglement of the theoretical uncertainty coming from modeling the nuclear potential and currents from that due to the approximations inherent in other many-body techniques. This is crucial for performing a comprehensive study of nuclear forces and properly assessing the theoretical uncertainty of the calculation. 

Light nuclei, i.e., those with $A \leq 12$, where $A$ is the number of nucleons, have proven to be an effective laboratory to test a variety of nuclear interaction models. In this realm, quantum Monte Carlo (QMC) methods have been extensively used to compute binding energies for both the ground- and the low-lying excited states at $\simeq 1\%$ accuracy level (see Ref.~\cite{carlson:2015} for a recent review).

The definition of the potential describing three-nucleon ($3N$) interactions is a central issue in nuclear theory. These forces are known to yield attractive contributions to the energy per particle of light nuclei. On the other hand, a repulsive contribution is needed for the stability of neutron stars against gravitational collapse and to reproduce the equilibrium properties of isospin-symmetric nuclear matter (SNM)~\cite{akmal:1998,carbone:2014,logoteta:2016}.

The most accurate phenomenological Hamiltonian for $A \leq 12$ nuclei comprises the Argonne $v_{18}$ (AV18)~\cite{wiringa:1995} two-nucleon ($N\!N$) potential and the Illinois-7 (IL7)~\cite{pieper:2001_prc,Pieper:2008} $3N$ potential. This provides a good description of the spectrum of nuclei up to $^{12}$C~\cite{pieper:2008_2} but yields a pathological equation of state of pure neutron matter~\cite{maris:2013}. On the other hand, when constraints on the $3N$ interaction are inferred from saturation properties of symmetric nuclear matter, the resulting predictions for neutron stars are compatible with astrophysical observations~\cite{gandolfi:2012,steiner:2012}. However $p$-shell light nuclei turn out to be underbound compared to experiment 
by about $0.25-0.75\,{\rm MeV}/A$~\cite{pieper:2001_prc}.

Elucidating the role of $3N$ forces in the region of medium-mass nuclei, such as $^{16}$O and $^{40}$Ca, is of paramount importance. Studying these two nuclei will help us to understand the mass region where the $3N$ contribution might already become repulsive. This aspect is strongly connected to the long-standing problem of the oxygen and calcium drip lines, which will be a major experimental focus of the Facility for Rare Isotope Beams~\cite{FRIB}. 

An accurate description of $^{16}$O, in particular its interaction with neutrinos, is also of immediate importance for the detection of supernova neutrinos~\cite{ankowski:2016}. The large water-Cherenkov detectors require  precise determination of their backgrounds, especially the one involving neutron knockout through neutral-current scattering of atmospheric neutrinos on $^{16}$O~\cite{ankowski:2013}. The computation of the electromagnetic responses of $^{16}$O using realistic nuclear interactions is a first step in this direction. In addition, studying the Coulomb sum rules of both $^{16}$O and $^{40}$Ca allows the investigation of putative in-medium modifications of the nucleon electromagnetic form factors~\cite{cloet:2016}.

Highly advanced nuclear many-body techniques, such as the coupled cluster method~\cite{hagen:2014}, the no-core shell model~\cite{barrett:2013}, the similarity renormalization group~\cite{hergert:2016}, and the self-consistent Green's function~\cite{dickhoff:2004}, have been successfully employed to study oxygen and calcium isotopes. In this work we use nuclear quantum Monte Carlo methods, which are capable of dealing with a wider range of momentum and energy, and allow the use of nuclear interactions characterized by high-momentum components. 

Standard quantum Monte Carlo techniques, namely variational Monte Carlo (VMC) and Green's function Monte Carlo (GFMC), work in the complete spin-isospin space, which grows exponentially with $A$~\cite{carlson:2015}. As a consequence, these methods are currently limited to $A \leq 12$ nuclei by available computational resources. Over the last two decades, the auxiliary field diffusion Monte Carlo (AFDMC) method~\cite{schmidt:1999,gandolfi:2014_prc,carlson:2015}, which uses Monte Carlo to also sample the spin-isospin degrees of freedom, has emerged as a more efficient algorithm for dealing with larger nuclear systems, but so far only for somewhat simplified interactions. Within cluster variational Monte Carlo (CVMC)~\cite{pieper:1990,pieper:1992}, expectation values are evaluated with a cluster expansion for the spin-isospin dependent correlations. The cluster expansion drastically reduces the computational effort necessary for the study of an $A$-body system, and it enables the study of medium-mass nuclei. Another approach based on a cluster expansion of nuclear correlations has been recently used to study the high-momentum components of nuclear wave functions (see~\cite{alvioli:2016} and references therein). This work, not based on Monte Carlo techniques, has been carried out employing two-body nuclear interactions only and limiting the cluster expansion to the leading order.

In this work we employ CVMC to perform variational calculations of three closed-shell nuclei, $^4$He, $^{16}$O, and $^{40}$Ca. We use as input a realistic phenomenological Hamiltonian, capable of describing the nucleon-nucleon data, both in scattering and bound states, with remarkable accuracy. The binding energy of the $3N$ system and the saturation density of isospin-symmetric nuclear matter are also well reproduced. We present results for the binding energy, charge radius, point density, single-nucleon momentum distribution, charge form factor, and Coulomb sum rule, fully taking into account the high-momentum components of the nuclear interaction. 

In Sec.~\ref{sec:hamiltonian} we briefly introduce the nuclear Hamiltonian and many-body wave functions used here. Section~\ref{sec:CVMC} is devoted to the description of the cluster variational Monte Carlo technique. In Sec.~\ref{sec:results} we present our results for $^4$He, $^{16}$O, and $^{40}$Ca. Finally, our conclusions are summarized in Sec.~\ref{sec:conclusions}.

\section{Nuclear Hamiltonian and wave functions}
\label{sec:hamiltonian}
Over a substantial range of energy and momenta, atomic nuclei can be described as collections of point-like particles of mass $m$, whose dynamics is dictated by a nonrelativistic Hamiltonian
\begin{align}
	H=-\frac{\hbar^2}{2m}\sum_i \nabla_{i}^2+\sum_{i<j}v_{ij}+\sum_{i<j<k} V_{ijk}\; .
\end{align}
Phenomenological $N\!N$ potentials include electromagnetic and one-pion-exchange terms at long range, and parametrize the intermediate- and short-distance region with phenomenological contributions that reproduce nucleon-nucleon elastic scattering data up to the pion-production threshold: 
\begin{align}
	v_{ij}=v_{ij}^{\gamma}+v_{ij}^\pi+v_{ij}^R \,.
\end{align}

A standard version in this class of potentials is the AV18~\cite{wiringa:1995} interaction. In AV18, the electromagnetic term $v_{ij}^{\gamma}$ includes one- and two-photon-exchange Coulomb interactions, vacuum polarization, Darwin-Foldy, and magnetic moment terms, with appropriate form factors that keep terms finite at $r_{ij}=0$, where $r_{ij} = |{\bm r}_i - {\bm r}_j|$ is the interparticle distance. The one-pion-exchange and phenomenological contributions can be written as a sum of 18 operators,
\begin{align}
	v_{ij}=\sum_{p=1}^{18} v^{p}(r_{ij})\,\mathcal O^{p}_{ij} \,.
\label{eq:NN}
\end{align}
The first six operators, corresponding to the static components of the $N\!N$ interaction, are
\begin{align}
	\mathcal O^{p=1,6}_{ij} = \big[\mathbbm{1}, \bm{\sigma}_{i}\cdot\bm{\sigma}_{j},S_{ij}\big]\otimes\big[\mathbbm{1},\bm{\tau}_{i}\cdot\bm{\tau}_{j}\big] ,
    \label{eq:v6}
\end{align}
where $\bm{\sigma}_{i}$ and $\bm{\tau}_{i}$ are Pauli matrices acting in spin and isospin space, respectively, and 
\begin{align}
	S_{ij}=3\,(\bm{\sigma}_{i}\cdot\hat{\bm r}_{ij}) (\bm{\sigma}_{j}\cdot\hat{\bm r}_{ij}) - (\bm{\sigma}_{i}\cdot\bm{\sigma}_{j}),
\end{align}
is the tensor operator. The operators $p=7,\ldots,14$ are associated with the non-static components of the $N\!N$ force. They have the form
\begin{align}
	\mathcal O^{p=7,14}_{ij} &= \big[\bm L\cdot\bm S,\bm L^2,\bm L^2\left(\bm{\sigma}_{i}\cdot\bm{\sigma}_{j}\right),(\bm L\cdot\bm S)^2\big] \nonumber \\
    &\,\otimes\big[\mathbbm{1},\bm{\tau}_{i}\cdot\bm{\tau}_{j}\big] ,
    \label{eq:v14}
\end{align}
where
\begin{align}
	\begin{aligned}
		\bm L &= \frac{1}{2i}\left(\bm r_i-\bm r_j\right)\times\left(\bm\nabla_i-\bm\nabla_j\right) , \\
	    \bm S &=\frac{1}{2}\left(\bm\sigma_i+\bm\sigma_j\right) ,
	\end{aligned}
\end{align}
are the relative angular momentum and the total spin of the pair $ij$, respectively. Overall, the first 14 operators of AV18 describe the charge-independent part of the $N\!N$ interaction. The last four operators account for small violations of isospin symmetry, and are grouped into charge-dependent $(p=15,17)$ and charge-symmetry breaking $(p=18)$ components,
\begin{align}
	\begin{aligned}
		\mathcal O^{p=15,17}_{ij} &= \big[\mathbbm{1}, \bm{\sigma}_{i}\cdot\bm{\sigma}_{j},S_{ij}\big]\otimes T_{ij} , \\
	    \mathcal O^{p=18}_{ij} &= \tau_{z_i}+\tau_{z_j} ,
	\end{aligned}
\end{align}
where $T_{ij}=3\,\tau_{z_i}\tau_{z_j}-\bm\tau_i\cdot\bm\tau_j$ is the isotensor operator.

AV18 fits the 1993 Nijmegen database~\cite{stoks:1993}, which includes $4301$ $N\!N$ scattering data up to $E_{\rm lab}=350\,\rm MeV$, with a $\chi^2/N_{data}\simeq 1.1$, as well as the deuteron binding energy and $nn$ scattering length. It is also found to be qualitatively good to much higher energies (up to $600\,\rm MeV$)~\cite{gandolfi:2014_epja}.

The inclusion of the $3N$ interaction $V_{ijk}$ is needed to explain the binding energies of the $3N$ systems and the saturation properties of SNM.  The derivation  of  $V_{ijk}$  was first discussed in the pioneering work of  Fujita and Miyazawa~\cite{fujita:1957}, who argued that its main contribution originates from the two-pion-exchange process in which the $N\!N$ interaction leads to the excitation of one of the participating nucleons to a (virtual) $\Delta$ resonance, which then decays by interacting with a third nucleon. 

In this work we use a phenomenological model of the $3N$ force, namely the Urbana IX (UIX) potential~\cite{pudliner:1995}, which is written as a sum of three contributions:
\begin{align}
	V_{ijk}=V_{ijk}^{2\pi,A}+V_{ijk}^{2\pi,C}+V_{ijk}^{R} \label{eq:v_ijk} \,.
\end{align}
The Fujita-Miyazawa anticommutator and commutator terms are
\begin{align}
	V_{ijk}^{2\pi,A}&=A_{2\pi}\sum_{cyc}\big\{X_{ij},X_{jk}\big\}\big\{\bm\tau_i\cdot\bm\tau_j,\bm\tau_j\cdot\bm\tau_k\big\} \label{eq:v_ijk_a} , \\
	V_{ijk}^{2\pi,C}&=\frac{A_{2\pi}}{4}\sum_{cyc}\big[X_{ij},X_{jk}\big]\big[\bm\tau_i\cdot\bm\tau_j,\bm\tau_j\cdot\bm\tau_k\big] \label{eq:v_ijk_c} ,
\end{align}
where $cyc$ denotes a cyclic sum over the three particle indexes and
\begin{align}
	X_{ij}&=Y_\pi(\mu_\pi r_{ij})\,\bm\sigma_i\cdot\bm\sigma_j+T_\pi(\mu_\pi r_{ij})\,S_{ij} , \\
    Y_\pi(x)&=\frac{e^{-x}}{x}\,\xi(r) , \label{eq:y_pi} \\
    T_\pi(x)&=\left(1+\frac{3}{x}+\frac{3}{x^2}\right)Y_\pi(x)\,\xi(r) \label{eq:t_pi} ,
\end{align}
with $\mu_\pi=m_\pi/\hbar c$ the pion mass, and $Y_\pi(x)$ and $T_\pi(x)$ the Yukawa and tensor Yukawa functions respectively, with cutoffs 
\begin{align}
	\xi(r)=1-e^{-c r^2} \label{eq:cutoff} \,.
\end{align}
The purely phenomenological repulsive term is given by
\begin{align}
	V_{ijk}^{R}=A_R\sum_{cyc}T^2_\pi(\mu_\pi r_{ij})\,T^2_\pi(\mu_\pi r_{ik}) \label{eq:v_ijk_r} \; .
\end{align}
The parameters $A_{2\pi}$ and $A_R$ are adjusted to reproduce the ground-state energy of the $3N$ systems and the SNM saturation density when used in conjunction with the AV18 $N\!N$ interaction.
The IL7 $3N$ potential also includes multi-pion-exchange components. The resulting AV18+IL7 Hamiltonian leads to predictions of $\simeq 100$ ground- and excited-state energies up to $A=12$ nuclei in very good agreement with the corresponding empirical values~\cite{carlson:2015}. However, when used to compute the neutron star matter equation of state, IL7 does not provide sufficient repulsion to guarantee the stability of observed stars against gravitational collapse~\cite{maris:2013}. We have therefore used the simpler UIX interaction in this study.

We note that the local $N\!N$ potentials recently derived within chiral perturbation theory~\cite{gezerlis:2013,gezerlis:2014,lynn:2014,piarulli:2015,piarulli:2016} are written in the same fashion as in \cref{eq:NN}. Because local versions of the chiral $3N$ potentials~\cite{tews:2016,lynn:2016,logoteta:2016} have spin-isospin structure analogous to that of UIX, the formalism developed in this paper can be readily applied to this class of interactions. 

Variational Monte Carlo exploits the stochastic Metropolis algorithm~\cite{metropolis:1953} to evaluate the expectation value of a given many-body operator using a suitably parametrized trial wave function $\Psi_V$. The nuclear potential introduces spin-isospin correlations into the nuclear wave function so the variational wave function should, to the extent possible, contain operator correlations of $v_{ij}$ and $V_{ijk}$. In the same spirit of Ref.~\cite{pieper:1992}, in this work we assume that a good variational wave function for the ground state of a closed-shell nucleus can be expressed as the product of two- and three-body correlation operators acting on a Jastrow wave function $\Psi_J$:
\begin{align}
	|\Psi_V\rangle&=\Bigg(1+\sum_{i<j<k}\,U_{ijk}\Bigg)\Bigg[\mathcal S\prod_{i<j}\,\Big(1+U^{2-6}_{ij}\Big)\Bigg] \nonumber \\
    & \quad\;\times \Bigg[1+\sum_{i<j}U^{7-8}_{ij}\Bigg]  |\Psi_J\rangle , \label{eq:psi_v} \\
    |\Psi_J\rangle&=\Bigg[\prod_{i<j}f_c(r_{ij})\Bigg]\mathcal{A}\,|\Phi\rangle \,. \label{eq:psi_j}
\end{align}

In the above equations, $U_{ij}$ and $U_{ijk}$ are correlations depending upon the spin and isospin of particles $ij$ and $ijk$, respectively. The $U^{2-6}_{ij}$ are static correlations (they contain no derivatives) while $U^{7-8}_{ij}$ are $\bm L\cdot\bm S$ correlations, fully defined following \cref{eq:uij}. The first term in the parentheses comes from the approximation of the independent triplet product of $(1+U_{ijk})$ to the linear term only. The symmetrization operator $\mathcal{S}$ is needed for the wave function to be fully antisymmetric, because $[U^{2-6}_{ij},U^{2-6}_{jk}]\neq 0$. To avoid multiple-order derivatives, the spin-orbit correlations $U^{7-8}_{ij}$ are done as a sum and act first on just the Jastrow wave function. In the Jastrow wave function, $f_c(r_{ij})$ denotes a central pair correlation function, $\mathcal A$ is the antisymmetrization operator, and $\Phi$ is an independent-particle wave function.

For doubly closed-shell nuclei, we can use a single product of four determinants $D_{\tau\sigma}$, one each for protons and neutrons, spin up and spin down, for $\Phi$:
\begin{align}
	|\Phi\rangle = \left\{D_{p\uparrow}\,D_{p\downarrow}\,D_{n\uparrow}\,D_{n\downarrow}\right\} ,
\end{align}
where each determinant contains $A/4$ nucleons. It follows that $\mathcal{A}|\Phi\rangle$ of \cref{eq:psi_j} is a sum over all the possible partitions of the $A$ nucleons into four groups of $A/4$ nucleons. 

Each determinant is constructed from single-particle radial wave functions
\begin{align}
	\phi_{nlm}(\bm{r})=R_{nl}(r)\,Y_{lm}(\theta,\varphi),\qquad \bm r=(r,\theta,\varphi) ,
\end{align}
calculated on the relative coordinates $\tilde{\bm r}_i$,
\begin{align}
	\tilde{\bm r}_i=\bm r_i -\bm R_{c.m.},\qquad\bm R_{c.m.}=\frac{1}{A}\sum_i \bm r_i ,
\end{align}
in order to make $\Phi$ translationally invariant. $Y_{lm}(\theta,\varphi)$ is the spherical harmonic. The radial wave functions $R_{nl}(r)$ are obtained from the bound-state solutions of the Woods-Saxon wine-bottle potential,
\begin{align}
	V(r)=V_s \Bigg[ \frac{1}{1+e^{(r-R_s)/a_s}}-\alpha_s e^{-(r/\rho_s)^2}\Bigg] , \label{eq:ws}
\end{align}
where the five parameters $V_S$, $R_s$, $a_s$, $\alpha_s$, and $\rho_s$ are determined variationally. 

As stated above, the two-body correlation operator $U_{ij}$ should reflect the spin-isospin structure of the underlying $N\!N$ potential. In this work we consider only the first eight spin-isospin operators, which capture the dominant features in the $N\!N$ phase shifts,
\begin{align}
	U_{ij}=\sum_{p=2}^8 \beta_p\,u_p(r_{ij})\,\mathcal O^{p}_{ij} ,
\label{eq:uij}
\end{align}
with $u_p(r_{ij})=f_p(r_{ij})/f_c(r_{ij})$. The radial correlation functions $f_{c,p}(r_{ij})$ are obtained by minimizing the two-body cluster contribution to the energy per particle of SNM at the Fermi momentum $k_F$. Euler-Lagrange (EL) equations are solved in a partial-wave $(S,T)$ basis for a quenched potential,
\begin{align}
	\bar{v}_{ij} = \sum_{p=1}^{14} \alpha_p\, v^p(r_{ij})\, \mathcal O^{p}_{ij} , 
\end{align}
by imposing the boundary conditions~\cite{lagaris:1981}:
\begin{align}
	\begin{aligned}
		f_c(r\geq d_1)&=1 ,  \\
		u_p(r\geq d_p)&=0 \; .
	\end{aligned}
\end{align}
In the present calculations we assume
\begin{align}
	\begin{aligned}
		\beta_{p=2-4,7-8}&=\beta_c, & \alpha_{p=1,9,13-14}=&1 , \\
		\beta_{p=5-6}    &=\beta_t, & \alpha_{p=2-8,10-12}=&\alpha , 
	\end{aligned}
\end{align}
and we consider three independent healing distances,
\begin{align}
	\begin{aligned}
		d_{p=1-4,7-8}&=d_{\rm S},d_{\rm P} ,\\
		d_{p=5-6}&=d_t ,
	\end{aligned}
\end{align}
where $d_{\rm S} \neq d_{\rm P}$ are used in order to differentiate $s$-wave ($^1S$ and $^3S-{^3D}$)  from $p$-wave ($^1P$ and $^3P-{^3F}$) channels, and the general relation $d_{\rm S}<d_{\rm P}<d_t$ should hold.  The functions $f_{c,p}(r_{ij})$ are projected from the solutions of the ($S,T$) partial-wave EL equations.  The pair correlation functions are thus fully specified by a total of seven variational parameters: $k_F$, $\alpha$, $\beta_c$, $\beta_t$, $d_{\rm S}$, $d_{\rm P}$, and $d_t$. 

In a many-body system it has been found advantageous to screen the spin- and isospin-dependent pair correlation functions when other particles are nearby~\cite{lomnitz:1981,pudliner:1997}. This can be achieved by multiplying $U_{ij}$ by three-body correlation factors,
\begin{align}
	U_{ij}\to\prod_{k\neq i,j} f_3(r_{ij};r_{ik},r_{jk})\,U_{ij} , \label{eq:ind}
\end{align}
where
\begin{align}
	\begin{aligned}
		&f_3(r_{ij};r_{ik},r_{jk})=1-t_1\left(\frac{r_{ij}}{R_{ijk}}\right)^{t_2}e^{-t_3 R_{ijk}} ,  \\
	    &R_{ijk}=r_{ij}+r_{ik}+r_{jk} \,.
	\end{aligned}
	\label{eq:f3}
\end{align}
The three parameters $t_1$, $t_2$, and $t_3$ are found variationally. 

Explicit triplet correlations significantly improve the variational energy for Hamiltonians including a $3N$ interaction. In this work we employed the form
\begin{align}
	U_{ijk}= \varepsilon_{2\pi,A}\,\tilde{V}^{2\pi,A}_{ijk}+\varepsilon_{R}\,\tilde{V}^{R}_{ijk} ,
\end{align}
where $\tilde{V}_{ijk}$ have the structures of \cref{eq:v_ijk_a,eq:v_ijk_r} but the two-particle distances are rescaled by a factor $\eta$, and two different constants $c_y$ and $c_t$ are used for the cutoff function $\xi(r)$ of \cref{eq:cutoff} used in \cref{eq:y_pi,eq:t_pi}. The triplet correlation functions are then given in terms of five variational parameters: $\varepsilon_{2\pi,A}$, $\varepsilon_R$, $\eta$, $c_y$, and $c_t$.

We did not include correlations arising from the commutator of \cref{eq:v_ijk_c} because it is significantly more computationally expensive to evaluate than the anticommutator of \cref{eq:v_ijk_a}. However, it has been shown that most of the correlations induced by the commutator can be effectively obtained by an appropriate choice of the coefficient $\varepsilon_{2\pi,A}$~\cite{pudliner:1997}.

\section{Cluster variational Monte Carlo}
\label{sec:CVMC}
In VMC, once the form for the trial wave function is assumed, one optimizes the variational parameters, typically by minimizing the expectation value and/or the variance of the total energy with respect to the variations of the parameters. The energy expectation value is given by
\begin{align}
	E_V=\frac{\langle\Psi_V|H|\Psi_V\rangle}{\langle\Psi_V|\Psi_V\rangle}\geq E_0,
\end{align}
and it is always greater than or equal to the ground-state energy with the same quantum numbers as $\Psi_V$. By minimizing $E_V$ the optimal $\Psi_V$ is obtained, and it is used to evaluate other quantities of interest. 

In general, for spin-isospin dependent interactions, the wave function is a sum of complex amplitudes for each spin-isospin state. The number of these components grows exponentially with the number of particles. This scaling can be mitigated by considering charge conservation and by assuming that the nucleus has good isospin $T$. However, for $A>12$ nuclei, quantum Monte Carlo calculations employing the complete many-body wave function currently represent a computational challenge~\cite{carlson:2015}.

One way to overcome the scaling problem and perform calculations for larger systems is to employ a cluster expansion scheme. The expectation value $\langle\Psi_V|H|\Psi_V\rangle$ as well as $\langle\Psi_V|\Psi_V\rangle$ can be expanded according to the number of nucleons connected by the spin-isospin correlations $U_{ij}$ and $U_{ijk}$. The resulting cluster expansion for the expectation value $E_V$, which is constructed according to  Ref.~\cite{pandharipande:1979}, has been used up to four-body cluster for the VMC study of $^{16}$O~\cite{pieper:1992} and $_{~\Lambda}^{17}$O~\cite{Usmani:1995} with earlier versions of the phenomenological $N\!N$+$3N$ potentials. In this work the calculations have been performed including up to five-body cluster contributions and considering closed-shell nuclei as large as $^{40}$Ca. The modern AV18 $N\!N$ potential plus the UIX $3N$ force has been employed.

\subsection{Cluster expansion}
The trial wave function of \cref{eq:psi_v} contains a large number of terms because there are many ways of partitioning $A$ nucleons into four groups of $A/4$ nucleons that preserve the antisymmetrization of $\Psi_V$. However, since $H$ is a symmetric operator, we can reduce the problem by considering a trial wave function $\Psi_R$ not fully antisymmetric,
\begin{align}
	|\Psi_R\rangle&=\Bigg(1+\sum_{i<j<k}\,U_{ijk}\Bigg)\Bigg[\mathcal S\prod_{i<j}\,\Big(1+U_{ij}^{2-6}\Big)\Bigg] \nonumber \\ 
    & \quad\; \times \Bigg[1+\sum_{i<j}U^{7-8}_{ij}\Bigg] |\Psi_J^R\rangle , \label{eq:psi_r} \\
    |\Psi_J^R\rangle&=\Bigg[\prod_{i<j}f_c(r_{ij})\Bigg]|\Phi\rangle ,
\end{align}
and by re-defining the energy expectation value as
\begin{align}
	E_V=\frac{\langle\Psi_V|H|\Psi_R\rangle}{\langle\Psi_V|\Psi_R\rangle}\,.\label{eq:e_v}
\end{align}
The cluster expansion adopted in this work is the one associated with expectation values of the form~(\ref{eq:e_v}). In the reference work~\cite{pieper:1992} this cluster expansion is referred to as ``CEA.''

Let us consider the expectation value of a symmetric one-body operator $\mathcal O_i$:
\begin{align}
	\frac{\langle\Psi_V|\sum_i\mathcal O_i|\Psi_R\rangle}{\langle\Psi_V|\Psi_R\rangle}=\frac{N}{D}=C \,. \label{eq:nond}
\end{align}
The numerator $N$ and denominator $D$ can be expanded as a sum of $n$-body contributions,
\begin{align}
	\!\!\!N&=\sum_i n_i+\sum_{i<j}n_{ij}+\!\!\sum_{i\ne j<k}\!n_{i,jk}+\!\!\sum_{i<j<k}\!\!n_{ijk}+\ldots ,  \label{eq:n}\\
    \!\!\!D&=1+\sum_{i<j}d_{ij}+\!\!\sum_{i<j<k}\!d_{ijk}+\!\!\sum_{\substack{i<j\ne k<l\\i<k}}\!\!d_{ij,kl}+\ldots \,. \label{eq:d}
\end{align}
Obviously extending the sums to $A$-body contributions gives the exact expectation value.
We define the generic expectation value $\langle X\rangle$, to be used for both $N$ and $D$ terms in \cref{eq:nond}, as
\begin{align}
	\langle X\rangle=\frac{\displaystyle\langle\Phi|\mathcal A \Bigg[\prod_{i<j}f_c(r_{ij})\Bigg] X \Bigg[\prod_{i<j}f_c(r_{ij})\Bigg]|\Phi\rangle}{\displaystyle\langle\Phi|\Bigg[\prod_{i<j}f_c(r_{ij})\Bigg]^2 |\Phi\rangle}\,. \label{eq:x}
\end{align}
The contributions $n_{ij\ldots}$ and $d_{ij\ldots}$ then take the following form:
\begin{align}
	n_i 	=&\left\langle\mathcal O_i\right\rangle , \nonumber \label{eq:n_i} \\
	n_{ij}	=& \left\langle\left(1+U_{ij}^\dagger\right)\left(\mathcal O_i+\mathcal O_j\right)\left(1+U_{ij}^{\phantom{\dagger}}\right)\right\rangle -n_i-n_j , \nonumber \\
	n_{i,jk}=&\left\langle\left(1+U_{jk}^\dagger\right)\mathcal O_i\left(1+U_{jk}^{\phantom{\dagger}}\right)\right\rangle - n_i , \nonumber \\
	n_{ijk}	=&\left\langle\Bigg[\mathcal S\prod_{cyc}\left(1+U_{ij}^\dagger\right)\Bigg]\left(1+U_{ijk}^\dagger\right)\left(\mathcal O_i+\mathcal O_j+\mathcal O_k\right)\right. \nonumber \\ 
		&\left.\left(1+U_{ijk}^{\phantom{\dagger}}\right)\Bigg[\mathcal S\prod_{cyc}\left(1+U_{ij}^{\phantom{\dagger}}\right)\Bigg]\right\rangle \nonumber \\ 
		&- \sum_{cyc}\left(n_{i,jk}+n_{ij}+n_i\right) , \\
	d_{ij}	&=\left\langle\left(1+U_{ij}^\dagger\right)\left(1+U_{ij}^{\phantom{\dagger}}\right)\right\rangle -1 \,. \label{eq:d_ij} 
\end{align}

The expansions (\ref{eq:n}) and (\ref{eq:d}) for $N$ and $D$ are divergent. On the other hand, a convergent expansion is achieved by considering the linked cluster expansion
\begin{align}
	C&=\sum_i c_i+\sum_{i<j}c_{ij}+\!\!\sum_{i\ne j<k}\!c_{i,jk}+\!\!\sum_{i<j<k}\!\!c_{ijk}+\ldots , \label{eq:c}
\end{align}
whose coefficients can be obtained from the equation $C\cdot D=N$ by equating terms containing the same number of particles,
\begin{align}
	\begin{aligned}
		c_i &= n_i ,  \\
	    c_{ij} &= \frac{n_{ij}-(c_i+c_j)\,d_{ij}}{1+d_{ij}} , \\	
	    c_{i,jk} &= \frac{n_{i,jk}-c_id_{jk}}{1+d_{jk}} ,  \\
	    c_{ijk} &= \!\frac{n_{ijk}-\!\displaystyle\sum_{cyc}\!\Big[c_id_{ijk}+(c_{ij}+c_{k,ij})(d_{ik}+d_{jk}+d_{ijk})\!\Big]}{1+\displaystyle\sum_{cyc}d_{ij}+d_{ijk}}
	\end{aligned}
\end{align}

The cluster expansion for the expectation value of two-body operators $\mathcal O_{ij}$ and three-body operators $\mathcal O_{ijk}$, such as $v_{ij}$ and $V_{ijk}$, resembles the one for the one-body operator $\mathcal O_i$. However, in the case of $\mathcal O_{ij}$, there are no one-body terms $n_i$, nor terms such as $n_{i,jk}$ in the numerator~(\ref{eq:n}). Therefore the cluster expansion (\ref{eq:c}) only contains terms of the kind $c_{ij},\,c_{ijk},\,c_{ij,kl},\,c_{ijkl},\,\ldots$. In a similar fashion, the cluster expansion for $\sum_{ijk}\mathcal O_{ijk}$ only comprises terms like $c_{ijk},\,c_{ijkl},\,c_{ijk,lm},\,c_{ijklm},\,\ldots$.

Terms such as $c_{i,jk}$ are referred to as semifactorizable. They are typically small because of the large cancellation between $n_{i,jk}$ and $c_i d_{jk}$, but they are finite. It is not necessary to treat them separately from the others. For example it is possible to define cluster contributions $\tilde{c}_{ijk}$ as the sum of all those that contain particles $ijk$ so that
\begin{align}	
	\tilde{c}_{ijk}=c_{ijk}+c_{i,jk}+c_{j,ik}+c_{k,ij} \,.
\end{align}
The corresponding $\tilde{n}_{ijk}$ can also be directly computed without separating their semifactorizable contributions. The total $n$-body cluster contribution $C_n$ is then obtained from the sum
\begin{align}
	C_n=\sum_{i_1<i_2<\ldots<i_n}\tilde{c}_{i_1 i_2 \ldots i_n} ,
\end{align}
and \cref{eq:c} can be simply rewritten as
\begin{align}
	C&=\sum_n C_n \,. \label{eq:c_full}
\end{align}

In the present work the cluster expansion is carried out up to five-body cluster, $n=5$. Since the operators in the expectation value $n_{ij\ldots l}$ or $d_{ij\ldots l}$ only contain the spin and isospin of particles $ij\ldots l$, the spin and isospin of the other particles are unchanged and can be ignored. If $ij\ldots l$ are in a single determinant $D_{\tau\sigma}$ in $|\Phi\rangle$, then only the term $\langle\Phi|$ in $\langle\Phi|\mathcal A$ contributes, and the rest can be ignored. If $i$ is in $D_{\tau^\prime\sigma^\prime}$ and $j\ldots l$ are in $D_{\tau\sigma}$ in $|\Phi\rangle$, then only the direct term $\langle\Phi|$ and those obtained by exchanging $i$ with $j\ldots l$ in $\langle\Phi|\mathcal A$ need to be considered. This implies a large reduction of the number of contributions to be calculated at each order, allowing for a full evaluation up to five-body cluster. 

All the expectation values $n_i,\,n_{ij},\,\ldots $ and $d_{ij},\,d_{ijk},\,\ldots$ are calculated up to four-body cluster. Five-body cluster contributions are instead sampled according to the probability
\begin{align}
	P(x)=\frac{1-P_{\min}}{1+e^{(x-b)/a}}+P_{\min} ,
\end{align}
where $x=\sum_{i<j}r_{ij}$, and typical values are $P_{min}=0.02$, $b=35\,\rm fm$, and $a=3.2\,\rm fm$. If $P(x)$ is larger than $\xi$, where $\xi$ is a random number in the interval $[0,1]$, then the five-body contribution is calculated. For $^{16}$O it has been verified that sampling  five-body cluster terms yields an energy expectation value that is compatible to the one obtained with the full five-body cluster calculation $(P_{\min}=1)$. In $^{16}$O the sampling procedure speeds up the evaluation of $E_V$ by a factor of 1.7 when using the $N\!N$ potential only, and by a factor of 2.2 when also $3N$ interactions are included. This is crucial for the calculation of $^{40}$Ca, in particular when using the full AV18+UIX potential. In $^{16}$O there are 4368 quintuplets, while in $^{40}$Ca there are 658008 quintuplets, making the full five-body cluster calculation extremely time demanding.

Further simplifications can be made by looking at the structure of the employed trial wave function. $|\Phi\rangle$ is a product of four determinants in which particle $(1:d)$, $(d+1:2d)$, $(2d+1:3d)$, and $(3d+1:4d)$, with $d=A/4$ are, respectively, $p\uparrow$, $p\downarrow$, $n\uparrow$ and $n\downarrow$. $\langle\Phi|\mathcal A$ is instead fully antisymmetric, so that when particle $i$ and $i^\prime$ belong to the same determinant, the following equivalences among expectation values apply:
\begin{align}
	\begin{aligned}
		n_{ij\ldots l}=n_{i^\prime j\ldots l} , \\
		d_{ij\ldots l}=d_{i^\prime j\ldots l} \,.
	\end{aligned}
\end{align}

By neglecting the effects of the Coulomb potential on the wave function, for the isospin-symmetric nuclei considered in this work it follows that, for instance, there are only four nonequivalent classes of $n_{ij}$ contributions:
\begin{align}
   \begin{aligned}
   	   n_{p\uparrow p\uparrow}&=n_{p\downarrow p\downarrow}=n_{n\uparrow n\uparrow}=n_{n\downarrow n\downarrow}, \\
       n_{p\uparrow p\downarrow}&=n_{n\uparrow n\downarrow} , \\
       n_{p\uparrow n\uparrow}&=n_{p\downarrow n\downarrow} , \\
       n_{p\uparrow n\downarrow}&=n_{p\downarrow n\uparrow} \,.
   \end{aligned}
\end{align}

We note that the employed cluster expansion treats exactly all the exchanges and central correlations among the $A$ nucleons. Every term in the cluster expansion (\ref{eq:n_i}) and (\ref{eq:d_ij}) contains the complete product of central correlations. In the conventional cluster expansions~\cite{pandharipande:1979}, one also expands in powers of $f_c^2(r)-1$ and this does not necessarily keep all the exchange terms. 

The current work includes the $\bm L\cdot\bm S$ correlations and $\bm L\cdot\bm S$, $\bm L^2$, and $(\bm L\cdot\bm S)^2$ potentials in all cluster expansion orders. Reference~\cite{pieper:1992} included these in only the two-body clusters, arguing that their total contribution is small. However we find a large, repulsive, three-body contribution from these potential terms.

Note that in the process of expanding the numerator and the denominator of the Hamiltonian's expectation value of \cref{eq:e_v}, the variational principle is not guaranteed to hold. However, since summing up to the $A$-body contribution gives the exact expectation value, the convergence of the cluster expansion itself will restore the validity of the variational principle. For this reason, during the optimization of the variational parameters, the convergence of the cluster expansion has been carefully checked for each of the analyzed cases.

\subsection{VMC sampling}
\label{sec:sampl}
The spatial integrals in \cref{eq:x} are evaluated using Metropolis Monte Carlo techniques~\cite{metropolis:1953}. The Metropolis method allows one to sample points in large-dimensional spaces according to a probability distribution $W(\bm R)$, where $\bm R=\{\bm r_1,\ldots,\bm r_A\}$. The algorithm generates a sequence of points ({\it random walk}) in the $3A$-dimensional space. This is achieved by a sequence of moves that can either be accepted or rejected depending upon the ratio of the function $W$ computed at the original and proposed points. 
According to the central limit theorem, the generic expectation value $\langle I\rangle$ can be written as
\begin{align}
	\langle I\rangle &=\frac{\int d\bm R\,W(\bm R)\,I(\bm R)}{\int d\bm R\,W(\bm R)} \nonumber \\ 
    & = \lim_{\mathcal N_c\to\infty}\frac{1}{\mathcal N_c}\sum_{i=1,\mathcal N_c}I(\bm R_i) ,
\end{align}
where $\mathcal N_c$ is the number of configurations $\bm R_i$ sampled with probability proportional to $W(\bm R)$. The Monte Carlo statistical error associated to $\langle I\rangle$ can be estimated with $\epsilon_I=\sqrt{\sigma_I/\mathcal N_c}$, where $\sigma_I$ is the variance of $I$.

The weight function $W(\bm R)$ must be positive definite and normalizable. The choice adopted in this work is to use the Jastrow part of the trial wave function $\Psi_R$
\begin{align}
	W(\bm R)=\Phi^*(\bm R)\Bigg[\prod_{i<j}f_c(r_{ij})\Bigg]^2\Phi(\bm R)\,F(\bm R) \,.
\end{align}
The expectation value $\langle X\rangle$ is
\begin{align}
	\langle X\rangle=\frac{\int d\bm R\,W(\bm R)\,\Phi^*(\bm R)\mathcal A\,X\,\Phi(\bm R)/\left[|\Phi(\bm R)|^2 F(\bm R) \right]}{\int d\bm R\,W(\bm R)/F(\bm R)} , \label{eq:x_sam}
\end{align}
and the function to evaluate at a sampled $\bm R_i$ is $\Phi^*(\bm R)\mathcal A\,X\,\Phi(\bm R)$ [with the normalization factor $|\Phi(\bm R)|^2 F(\bm R)$], where the spin-isospin summations are implicit. In the present case $\Phi(\bm R)$ is real, so that $\Phi^*(\bm R)=\Phi(\bm R)$. 

The factor $F(\bm R)$ is introduced in the weight function $W(\bm R)$ in order to prevent the quantity $\Phi^*(\bm R)\mathcal A\,X\,\Phi(\bm R)/|\Phi(\bm R)|^2$ from becoming very large. It is chosen so that $\Phi^*(\bm R)\mathcal A\,X\,\Phi(\bm R)/\!\left[|\Phi(\bm R)|^2 F(\rm R)\right]$ is finite at all $\bm R$. All the exchanges that contribute to $\Phi^*(\bm R)\mathcal A\,X\,\Phi(\bm R)$ are included in $|\Phi(\bm R)|^2 F(\bm R)$ so that
\begin{align}
	|\Phi(\bm R)|^2F(\bm R) =&\, |\Phi(\bm R)|^2 \nonumber \\
    & +\sum_{i<j}\omega(r_{ij})|\mathcal P_{ij}\Phi(\bm R)|^2 \nonumber \\ 
    & +\sum_{i<j<k}\omega(r_{ij})\,\omega(r_{jk})\,\omega(r_{ik}) \nonumber \\ 
    & \quad\;\;\times \Big[|\mathcal P_{ij}\mathcal P_{ik}\Phi(\bm R)|^2+|\mathcal P_{ik}\mathcal P_{ij}\Phi(\bm R)|^2\Big] \nonumber \\ 
    & +\cdots , \label{eq:F}
\end{align}
where $\mathcal P_{ij}$ is the exchange operator acting on particles $i$ and $j$. In the present work contributions up to four-body exchanges are considered in \cref{eq:F}. The function $\omega(r)$ is chosen to be proportional to the sum of the squares of $u_p(r)$, since the exchange of particles $i$ and $j$ with different spin-isospin states in $\Phi(\bm R)$ must be accompanied by a $v_{ij}$, $U_{ij}$, $V_{ijk}$, or $U_{ijk}$. The use of the importance function $F(\bm R)$ drastically reduces the variance on the expectation values. For instance, in $^{16}$O the same statistical error for the energy expectation value can be achieved by using just half of the configurations when $F(\bm R)\neq 1$.

The $\Phi(\bm R)\mathcal A\,X\,\Phi(\bm R)$ for a given cluster is calculated with methods developed for few-body systems~\cite{wiringa:1991,lomnitz:1981}. The terms in $\Phi(\bm R)\mathcal A$ that can contribute are summed, and $\Phi(\bm R)\mathcal A$ is represented as a vector whose components give the amplitudes of the spin-isospin states of the nucleons in the cluster. The corresponding vector representing $\Phi(\bm R)$ has only one nonzero component since all particles have definite values of $\tau_z$ and $\sigma_z$ in $\Phi(\bm R)$. The $v_{ij}$, $V_{ijk}$, $U_{ij}$, and $U_{ijk}$ operate on these vectors as discussed in Refs.~\cite{wiringa:1991,lomnitz:1981}. The expectation values of the kinetic energy operators are obtained by computing $\Psi_R$ at slightly shifted positions and using finite differences to evaluate terms in $\nabla^2\Psi_R$.

Due to the tremendous increase in computer power of the last decades, many of the approximations implemented in the reference work~\cite{pieper:1992} are no longer necessary. For instance, in the current calculations the three-body correlation operators $U_{ijk}$ act last in $\Psi_R$ and $\Psi_V$ of \cref{eq:psi_v,eq:psi_r}, as in the original formulation of the trial wave function. In Ref.~\cite{pieper:1992}, because of the computational limitations of the time, $\Psi_R$ and $\Psi_V$ were approximated acting first with the $U_{ijk}$  on the sparse vectors representing $\Phi(\bm R)$ and $\mathcal A \Phi(\bm R)$, and then operating with the two-body correlations $U_{ij}$. The latter are now implemented in all orders, including the spin-orbit correlations that were previously calculated at the two-body level only. 

Moreover, the calculation of the contribution of the kinetic energy, $N\!N$, and $3N$ potential operators is fully carried out at each order of the cluster expansion. For the five-body cluster, all the one-, two-, and three-body operators are evaluated, although their contributions to $C_5$ are sampled as previously discussed.

\subsection{Optimization}
The trial wave functions of \cref{eq:psi_r,eq:psi_v} contain a total of 15 variational parameters when only two-body correlations are considered, and up to 20 parameters if three-body correlations are also included. In order to perform the minimization of the energy expectation value with respect to these sets of parameters, we used the NLopt optimization tool, as recently done in other standard VMC calculations~\cite{piarulli:2016}.

NLopt is a free/open-source library for nonlinear optimization developed at the Massachusetts Institute of Technology~\cite{johnson}. It provides a common interface for a number of different free optimization routines available online as well as original implementations of various other algorithms, including both global and local optimization algorithms, both derivative-free and user-supplied gradients algorithms, and algorithms for unconstrained optimization, bound-constrained optimization, and general nonlinear inequality/equality constraints.

In this work we implemented different local derivative-free algorithms, and in particular we made extensive use of the COBYLA (constrained optimization bY linear approximations)~\cite{powell:1994,*powell:1998} and Nelder-Mead simplex~\cite{nelder:1965,*box:1965,*richardson:1972} algorithms. It has been observed that both algorithms perform well in the case of $^4$He~\cite{piarulli:2016}. For heavier systems, Nelder-Mead simplex seems instead to be the optimal algorithm, providing better convergence and reliability of the minimization search. This is probably related to the fact that the minimization was done using correlated energy differences~\cite{wiringa:1991} for $A=4$ but not for the larger nuclei. 

For both $^4$He and $^{16}$O the energy minimization was carried out in the full parameter space, with the energy expectation value calculated up to the highest cluster contribution for the system under study. In order to reduce the computational cost of the optimization process, the spin-orbit correlations are turned off during the variational search. However, once the optimal set of parameters is found, the full two-body correlations of \cref{eq:uij} are employed in the calculation of the expectation values. In the case of $^{40}$Ca, the computation of five-body cluster contributions to the total energy is quite demanding, even with the sampling procedure. Each CVMC run for $A=40$ requires approximately two hours on 18 32-core Intel Haswell 2.3~GHz nodes to obtain a statistical error of $\simeq 0.5\,{\rm MeV}/A$ for the energy of the full $N\!N$ plus $3N$ Hamiltonian. The variational search over the entire 20-dimensional parameter space would have required at least $\sim 140$~h on the same hardware configuration, i.e., more than $80\cdot10^3$ CPU hours. 

Relying on the observation that short-range correlations for medium-heavy systems should be independent of $A$, the energy minimization for $^{40}$Ca has been carried out in a subset of the parameter space. When using AV18 only, the optimal parameters for the two-body correlations found in $^{16}$O for the same potential were employed, as were the wine-bottle coefficients and the induced three-body correlations of \cref{eq:ind}. The variational search was performed for only the three parameters defining the Wood-Saxon potential of \cref{eq:ws}. When the $3N$ potential is also included, from the best set of parameters for $^{16}$O with the same interaction, we minimized over the three parameters of the Wood-Saxon potential and over two of the five parameters of the three-body correlations. For the latter, $c_y$ and $c_t$ appear to be the most effective to produce appreciable changes in the total energy. All the parameters for the systems under study for both AV18 and AV18+UIX are listed the Appendix.

\section{Results}
\label{sec:results}
The expectation values of all observables are calculated for each nucleus by summing all cluster contributions up to five-body cluster (four-body in the case of $^4$He). For $^{16}$O and $^{40}$Ca, the full expansion should consider contributions up to 16- and 40-body clusters, respectively. Under the observation that the ratio between the last successive cluster contributions is small and approximately constant, we can estimate the cluster contribution $C_{6-A}$ by assuming uniform convergence, i.e., by using the relation
\begin{align}
	\frac{C_{k+1}}{C_k}=\frac{C_k}{C_{k-1}} \,.\label{eq:Crat}
\end{align}
Given the cluster expansion at order $k$, the total extrapolated result is obtained by summing over all the cluster contributions, including the extrapolated ones
\begin{align}
	C_{\rm ext}=\sum_{n=1}^{k}C_n + \sum_{n=k+1}^{A}C_n^{\rm ext} \,. \label{eq:Cext1}
\end{align}
Under the assumption of uniform convergence, the $C_n^{\rm ext}$ form a geometric progression, and we can then recast the total $C_{\rm ext}$ using the sum of the geometric series
\begin{align}
	C_{\rm ext}=\sum_{n=1}^{k-2}C_n+C_{k-1}\,\frac{1}{1-x},\qquad x=\frac{C_{k}}{C_{k-1}} \,. \label{eq:Cext2}
\end{align}
Note that in the employed cluster expansion, successive cluster contributions $C_{k}$ and $C_{k-1}$ have decreasing magnitude and opposite sign, so that $|x|<1$.

\Cref{eq:Cext1,eq:Cext2} give consistent results for all the observables under study. In the following, unless otherwise specified, we will report results for $^{16}$O and $^{40}$Ca using the extrapolation of \cref{eq:Cext1} for contributions above the five-body cluster. Errors on $\sum_{n=k+1}^{A}C_n^{\rm ext}$ are estimated by propagating the CVMC statistical errors from the previous cluster contributions.

\subsection{Energies, radii, and densities}

\begin{table*}[ht]
\centering
\caption{Cluster contributions to the energy per nucleon and point radius in $^4$He when using the AV18 potential. Energies are in MeV/$A$ and radii in fm$^2$. Averages are calculated using $\mathcal N_c=2\times10^6$ configurations. In this and the following tables, the Monte Carlo statistical errors of the last digits are given in parentheses. A (0) indicates an error of less than 5 in the following digit.}
\label{tab:he4_av18}
\begin{tabular}{l c c c c c }
\hline\hline
observable        & 1b         & 2b          & 3b         & 4b          & sum         \\
\hline                                                                                  
$T$               & $12.22(2)$ & $12.51(1) $ & $-1.12(1)$ & $0.23(1)  $ & $23.84(3) $ \\ [0.2cm]
$v_{ij}^{1-6}$    & $        $ & $-30.07(3)$ & $0.52(1) $ & $-0.22(1) $ & $-29.77(3)$ \\
$v_{ij}^{7-14}$   & $        $ & $-1.00(0) $ & $1.07(0) $ & $-0.08(0) $ & $-0.01(0) $ \\
$v_{ij}^{\gamma}$ & $        $ & $0.20(0)  $ & $0.01(0) $ & $0.00(0)  $ & $0.21(0)  $ \\
$v_{ij}$          & $        $ & $-30.86(3)$ & $1.58(1) $ & $-0.29(1) $ & $-29.57(3)$ \\ [0.2cm]
$T+v_{ij}$        & $12.22(2)$ & $-18.35(2)$ & $0.47(1) $ & $-0.06(0) $ & $-5.73(1) $ \\ [0.2cm]
$r_{\rm pt}^2$    & $2.813(2)$ & $-0.568(1)$ & $0.046(0)$ & $-0.003(0)$ & $2.289(1) $ \\
\hline\hline
\end{tabular}
\end{table*}

\begin{table*}[ht]
\centering
\caption{Cluster contributions to the energy per nucleon and point radius in $^4$He when using the AV18+UIX potential. Energies are in MeV/$A$ and radii in fm$^2$. Averages are calculated using $\mathcal N_c=2\times10^6$ configurations.}
\label{tab:he4_av18+uix}
\begin{tabular}{l c c c c c }
\hline\hline
observable         & 1b         & 2b          & 3b         & 4b          & sum         \\
\hline                                                                                   
$T$                & $13.12(2)$ & $15.55(2) $ & $-2.30(1)$ & $0.44(1)  $ & $26.80(3) $ \\ [0.2cm]
$v_{ij}^{1-6}$     & $        $ & $-33.74(3)$ & $1.45(1) $ & $-0.30(1) $ & $-32.60(3)$ \\
$v_{ij}^{7-14}$    & $        $ & $-1.31(0) $ & $1.66(0) $ & $-0.19(0) $ & $0.17(3)  $ \\
$v_{ij}^{\gamma}$  & $        $ & $0.21(0)  $ & $0.01(0) $ & $0.00(0)  $ & $0.22(0)  $ \\
$v_{ij}$           & $        $ & $-34.83(3)$ & $3.10(2) $ & $-0.48(1) $ & $-32.21(3)$ \\ [0.2cm]
$V_{ijk}^{2\pi,A}$ & $        $ & $         $ & $-1.67(0)$ & $0.04(0)  $ & $-1.64(0) $ \\
$V_{ijk}^{2\pi,C}$ & $        $ & $         $ & $-1.02(0)$ & $0.04(0)  $ & $-0.98(0) $ \\
$V_{ijk}^{R}$      & $        $ & $         $ & $1.36(0) $ & $-0.03(0) $ & $1.33(0)  $ \\
$V_{ijk}$          & $        $ & $         $ & $-1.34(0)$ & $0.05(0)  $ & $-1.29(0) $ \\ [0.2cm]
$T+v_{ij}$         & $13.12(2)$ & $-19.29(2)$ & $0.80(0) $ & $-0.04(1) $ & $-5.41(1) $ \\
$T+v_{ij}+V_{ijk}$ & $13.12(2)$ & $-19.29(2)$ & $-0.54(1)$ & $0.01(1)  $ & $-6.70(1) $ \\ [0.2cm]
$r_{\rm pt}^2$     & $2.646(2)$ & $-0.600(1)$ & $0.070(0)$ & $-0.005(0)$ & $2.111(1) $ \\
\hline\hline
\end{tabular}
\end{table*}

The contributions of the cluster expansion to the kinetic energy $T$, to the $N\!N$ and $3N$ potentials, and to the point radius are listed in \cref{tab:he4_av18,tab:he4_av18+uix} for $^4$He, in \cref{tab:o16_av18,tab:o16_av18+uix} for $^{16}$O, and in \cref{tab:ca40_av18,tab:ca40_av18+uix} for $^{40}$Ca. The expectation value of $v_{ij}^{15-18}$ is zero for all the systems under study, as we are assuming pure $T=0$ ground states. Since one-, two-, and three-body operators exhibit different convergence patterns in the cluster expansion, for $A>4$ the total energy is estimated as the sum of the extrapolated results for $T$, $v_{ij}$, and $V_{ijk}$, and it is italicized in the tables. 

Let us first consider the $^4$He nucleus. By comparing the energy per nucleon obtained with AV18 and AV18+UIX Hamiltonians, both reported in \cref{tab:he4}, it is apparent that the $3N$ force gives overall $\simeq 1\,{\rm MeV}/A$ more binding. This result is consistent with VMC and GFMC calculations~\cite{pieper:2001_prc,pieper:2001_arnps} for the same interactions. It is interesting to note that the best wave function for the full Hamiltonian including UIX sacrifices $\simeq 0.3\,{\rm MeV}/A$ from the $T+v_{ij}$ contribution which is made up by increasing the attraction from UIX.

The more sophisticated two-body correlations employed in the VMC wave function for $s$-shell nuclei yield $\simeq 0.2\,{\rm MeV}/A$ additional binding compared to the CVMC results. Nevertheless, the CVMC energies are within $5\%$ from the GFMC values, and charge radii are remarkably close for all the three quantum Monte Carlo methods. This corroborates the accuracy of the wave functions employed in this work to describe the ground state of closed-shell nuclei, which, combined with the cluster expansion technique, allows reliable variational calculations for nuclei as heavy as $^{40}$Ca.

\begin{table*}[ht]
\centering
\caption{Cluster contributions to the energy per nucleon and point radius in $^{16}$O when using the AV18 potential. Energies are in MeV/$A$ and radii in fm$^2$. Averages are calculated using $\mathcal N_c=6\times10^5$ configurations.}
\label{tab:o16_av18}
\begin{tabular}{l c c c c c c c}
\hline\hline
observable        & 1b         & 2b          & 3b         & 4b          & 5b          & 6-16b       & sum             \\
\hline                        
$T$               & $19.78(2)$ & $16.67(2) $ & $-6.13(3)$ & $2.24(3)  $ & $-0.44(3) $ & $0.07(1)  $ & $32.19(5) $     \\ [0.2cm]
$v_{ij}^{1-6}$    & $        $ & $-48.30(4)$ & $10.69(3)$ & $-2.42(4) $ & $-0.59(5) $ & $-0.19(2) $ & $-40.81(6)$     \\
$v_{ij}^{7-14}$   & $        $ & $-0.46(0) $ & $3.48(1) $ & $-1.70(1) $ & $0.69(1)  $ & $-0.20(2) $ & $1.82(3)  $     \\
$v_{ij}^{\gamma}$ & $        $ & $0.91(0)  $ & $0.03(0) $ & $-0.02(0) $ & $0.00(0)  $ & $0.00(0)  $ & $0.91(0)  $     \\
$v_{ij}$          & $        $ & $-47.85(4)$ & $14.20(3)$ & $-4.15(4) $ & $0.10(5)  $ & $0.00(0)  $ & $-37.70(3)$     \\ [0.2cm]
$T+v_{ij}$        & $19.78(2)$ & $-31.18(2)$ & $8.07(2) $ & $-1.91(3) $ & $-0.34(3) $ & ---         & $\it -5.51(2) $ \\ [0.2cm]
$r_{\rm pt}^2$    & $6.235(2)$ & $-0.672(2)$ & $0.277(1)$ & $-0.074(1)$ & $-0.010(1)$ & $-0.002(0)$ & $5.754(3) $     \\
\hline\hline
\end{tabular}
\end{table*}

\begin{table*}[ht]
\centering
\caption{Cluster contributions to the energy per nucleon and point radius in $^{16}$O when using the AV18+UIX potential. Energies are in MeV/$A$ and radii in fm$^2$. Averages are calculated using $\mathcal N_c=6\times10^5$ configurations.}
\label{tab:o16_av18+uix}
\begin{tabular}{l c c c c c c c}
\hline\hline
observable         & 1b         & 2b          & 3b         & 4b          & 5b          & 6-16b       & sum             \\
\hline    
$T$                & $16.29(2)$ & $16.34(2) $ & $-3.79(2)$ & $0.90(2)  $ & $-0.05(3) $ & $0.00(0)  $ & $29.70(4) $     \\ [0.2cm]
$v_{ij}^{1-6}$     & $        $ & $-41.90(3)$ & $6.61(2) $ & $-0.95(4) $ & $-0.36(3) $ & $-0.22(4) $ & $-36.83(8)$     \\
$v_{ij}^{7-14}$    & $        $ & $-0.82(0) $ & $2.93(1) $ & $-1.01(1) $ & $0.27(1)  $ & $-0.06(1) $ & $1.31(2)  $     \\
$v_{ij}^{\gamma}$  & $        $ & $0.83(0)  $ & $0.03(0) $ & $-0.02(0) $ & $0.00(0)  $ & $0.00(0)  $ & $0.84(0)  $     \\
$v_{ij}$           & $        $ & $-41.89(3)$ & $9.57(2) $ & $-1.98(4) $ & $-0.10(4) $ & $-0.01(0) $ & $-34.41(2)$     \\ [0.2cm]
$V_{ijk}^{2\pi,A}$ & $        $ & $         $ & $-2.49(1)$ & $1.12(0)  $ & $-0.29(1) $ & $0.06(1)  $ & $-1.59(2) $     \\
$V_{ijk}^{2\pi,C}$ & $        $ & $         $ & $-1.69(0)$ & $0.79(0)  $ & $-0.26(0) $ & $0.06(0)  $ & $-1.10(0) $     \\
$V_{ijk}^{R}$      & $        $ & $         $ & $2.83(1) $ & $-0.70(0) $ & $0.12(1)  $ & $-0.02(0) $ & $2.23(1)  $     \\
$V_{ijk}$          & $        $ & $         $ & $-1.35(1)$ & $1.21(0)  $ & $-0.43(1) $ & $0.11(1)  $ & $-0.45(2) $     \\ [0.2cm]
$T+v_{ij}$         & $16.29(2)$ & $-25.54(2)$ & $5.78(2) $ & $-1.08(2) $ & $-0.14(2) $ & ---         & $\it -4.70(2) $ \\
$T+v_{ij}+V_{ijk}$ & $16.29(2)$ & $-25.54(2)$ & $4.43(2) $ & $0.13(2)  $ & $-0.57(2) $ & ---         & $\it -5.15(2) $ \\ [0.2cm]
$r_{\rm pt}^2$     & $7.353(3)$ & $-0.680(2)$ & $0.217(1)$ & $-0.035(2)$ & $-0.007(1)$ & $-0.002(0)$ & $6.846(3) $     \\
\hline\hline
\end{tabular}
\end{table*}

The total energies of $^{16}$O and $^{40}$Ca for both AV18 and AV18+UIX are reported in \cref{tab:o16,tab:ca40}, respectively. Our variational calculations show that the AV18+UIX Hamiltonian underbinds both $^{16}$O and $^{40}$Ca, by $2.83(3)\,{\rm MeV}/A$ and $3.63(10)\,{\rm MeV}/A$, respectively. The results obtained for $^{40}$Ca are consistent with variational calculations for SNM performed with the same interaction~\cite{akmal:1998}, which yield $-11.85\,{\rm MeV}/A$, to be compared to the empirical value of $\simeq -16\,{\rm MeV}/A$. This underbinding can be only partly ascribed to deficiencies of the variational wave function, which has proven to be accurate for describing infinite matter properties~\cite{lovato:2011}. To gauge the accuracy of the CVMC wave function in describing closed-shell nuclei, we performed a benchmark calculation with AFDMC using the AV6$^\prime$ potential. This is a reprojection of the full AV18 onto the first six operators that preserves the deuteron binding energy and many of the properties of elastic $N\!N$ scattering~\cite{wiringa:2002}. To obtain $^{16}$O energies that are bound against $\alpha$-particle break up, the Coulomb interaction was omitted. The results listed in \cref{tab:av6p} show a $\simeq 0.25\,\rm MeV/A$ and a $\simeq0.45\,\rm MeV/A$ energy difference in $^4$He and $^{16}$O respectively between CVMC and AFDMC results. This is expected for a variational versus a diffusion Monte Carlo calculation. Charge radii are instead compatible between the two methods, confirming the quality of the employed wave function. Therefore, a large fraction of the missing binding in $^{16}$O and $^{40}$Ca is due to limitations of the AV18+UIX Hamiltonian. Note that these results show significantly less binding per nucleon for both $^{16}$O and $^{40}$Ca than for $^4$He, i.e., they predict that $^{16}$O and $^{40}$Ca would break apart into $^4$He nuclei. However, \cref{tab:av6p} indicates that $^{16}$O is stable against breakup with the AV6$^\prime$ interaction if the Coulomb interaction is omitted.

For both $^{16}$O and $^{40}$Ca the expectation value of $V_{ijk}^{2\pi}$ is negative, and that of $V_{ijk}^R$ is positive, leading to an overall attractive contribution of the $3N$ force, as for $^4$He. However, by comparing the total energies for AV18 and AV18+UIX, it turns out that $^{16}$O and $^{40}$Ca are less bound when the $3N$ force is included. This is particularly evident in $^{40}$Ca, where the UIX potential reduces the binding energy of $\simeq1\,{\rm MeV}/A$. This is somewhat consistent with the fact that the UIX force is repulsive in SNM. Finally, it is interesting to notice how, within a variational approach, the change in the behavior of the employed $3N$ force---from attractive to repulsive---is already manifest in relatively small nuclear systems, like $^{16}$O. 

\begin{table*}[htb]
\centering
\caption{Cluster contributions to the energy per nucleon and point radius in $^{40}$Ca when using the AV18 potential. Energies are in MeV/$A$ and radii in fm$^2$. Averages are calculated using $\mathcal N_c=5\times10^5$ configurations.}
\label{tab:ca40_av18}
\begin{tabular}{l c c c c c c c c}
\hline\hline
observable        & 1b          & 2b          & 3b         & 4b          & 5b          & 6-40b      & sum             \\
\hline                                                                                                   
$T$               & $20.80(1) $ & $18.00(1) $ & $-8.60(3)$ & $4.04(3)  $ & $-1.22(10)$ & $0.28(6) $ & $32.29(15) $    \\ [0.2cm]
$v_{ij}^{1-6}$    & $         $ & $-54.52(3)$ & $15.80(2)$ & $-5.09(6) $ & $0.05(9)  $ & $0.00(0) $ & $-43.76(8) $    \\
$v_{ij}^{7-14}$   & $         $ & $-0.16(0) $ & $4.45(1) $ & $-2.78(1) $ & $1.47(2)  $ & $-0.51(5)$ & $2.47(7)   $    \\
$v_{ij}^{\gamma}$ & $         $ & $1.86(0)  $ & $0.07(0) $ & $-0.08(0) $ & $0.00(0)  $ & $0.00(0) $ & $1.85(0)   $    \\
$v_{ij}$          & $         $ & $-52.81(3)$ & $20.31(2)$ & $-7.95(6) $ & $1.52(10) $ & $-0.24(3)$ & $-39.17(11)$    \\ [0.2cm]
$T+v_{ij}$        & $20.80(2) $ & $-34.81(2)$ & $11.71(3)$ & $-3.91(5) $ & $0.30(9)  $ & ---        & $\it -5.88(10)$ \\ [0.2cm]
$r_{\rm pt}^2$    & $11.204(3)$ & $-0.920(4)$ & $0.506(3)$ & $-0.188(5)$ & $0.006(7) $ & $0.000(0)$ & $10.609(7) $    \\
\hline\hline
\end{tabular}
\end{table*}

\begin{table*}[htb]
\centering
\caption{Cluster contributions to the energy per nucleon and point radius in $^{40}$Ca when using the AV18+UIX potential. Energies are in MeV/$A$ and radii in fm$^2$. Averages are calculated using $\mathcal N_c=5\times10^5$ configurations.}
\label{tab:ca40_av18+uix}
\begin{tabular}{l c c c c c c c}
\hline\hline
observable         & 1b          & 2b          & 3b         & 4b          & 5b          & 6-40b       & sum             \\
\hline
$T$                & $17.35(1) $ & $17.86(1) $ & $-5.65(2)$ & $1.83(3)  $ & $-0.54(9) $ & $0.12(5)  $ & $30.97(14) $    \\ [0.2cm]
$v_{ij}^{1-6}$     & $         $ & $-47.20(3)$ & $10.20(2)$ & $-2.33(5) $ & $0.09(8)  $ & $0.00(0)  $ & $-39.24(7) $    \\
$v_{ij}^{7-14}$    & $         $ & $-0.67(0) $ & $3.68(1) $ & $-1.60(1) $ & $0.52(2)  $ & $-0.13(1) $ & $1.80(3)   $    \\
$v_{ij}^{\gamma}$  & $         $ & $1.72(0)  $ & $0.06(0) $ & $-0.05(0) $ & $0.00(0)  $ & $0.00(0)  $ & $1.73(0)   $    \\
$v_{ij}$           & $         $ & $-46.15(2)$ & $13.94(2)$ & $-3.99(5) $ & $0.62(8)  $ & $-0.08(2) $ & $-35.66(9) $    \\ [0.2cm]
$V_{ijk}^{2\pi,A}$ & $         $ & $         $ & $-2.98(0)$ & $1.86(0)  $ & $-0.71(1) $ & $0.20(1)  $ & $-1.63(2)  $    \\
$V_{ijk}^{2\pi,C}$ & $         $ & $         $ & $-2.10(0)$ & $1.33(0)  $ & $-0.61(1) $ & $0.19(1)  $ & $-1.19(2)  $    \\
$V_{ijk}^{R}$      & $         $ & $         $ & $3.59(1) $ & $-1.25(0) $ & $0.28(1)  $ & $-0.05(0) $ & $2.57(1)   $    \\
$V_{ijk}$          & $         $ & $         $ & $-1.48(0)$ & $1.93(1)  $ & $-1.05(1) $ & $0.37(3)  $ & $-0.23(4)  $    \\ [0.2cm]
$T+v_{ij}$         & $17.35(1) $ & $-28.29(2)$ & $8.30(2) $ & $-2.16(3) $ & $0.08(8)  $ & ---         & $\it -4.69(9)$  \\
$T+v_{ij}+V_{ijk}$ & $17.35(1) $ & $-28.29(2)$ & $6.81(2) $ & $-0.23(3) $ & $-0.97(8) $ & ---         & $\it -4.92(10)$ \\ [0.2cm]
$r_{\rm pt}^2$     & $13.025(3)$ & $-0.903(4)$ & $0.393(2)$ & $-0.091(5)$ & $-0.022(7)$ & $-0.007(4)$ & $12.394(11)$    \\
\hline\hline
\end{tabular}
\end{table*}

\begin{table*}[htb]
\centering
\caption[]{Total energies (in MeV/$A$) and charge radii (in fm) in $^4$He for different potentials. VMC and GFMC results are taken from Refs.~\cite{pieper:2001_prc,pieper:2001_arnps}.}
\label{tab:he4}
\begin{tabular}{l l c c c c}
\hline\hline
obs & potential & CVMC & VMC & GFMC & exp \\
\hline
\multirow{2}{*}{$\left\langle E\right\rangle$} & AV18 & $-5.73(1)$ & $-5.93(1)$ & $-6.02(1)$ & \multirow{2}{*}{$-7.07$} \\
& AV18+UIX & $-6.70(1)$ & $-6.95(1)$ & $-7.08(1)$ & \\ [0.2cm] 
\multirow{2}{*}{$\sqrt{\left\langle r_{\rm ch}^2\right\rangle}$} & AV18 & $1.725(3)$ & $1.734(3)$ & --- & \multirow{2}{*}{$1.676(3)$} \\ 
& AV18+UIX & $1.673(3)$ & $1.665(3)$ & $1.661(3)$ & \\
\hline\hline
\end{tabular}
\end{table*}

Three-nucleon forces significantly affect quantities other than the energy, such as point radii and point densities. The latter are related to the charge density, which can be extracted from electron-nucleus scattering data, but they are not observables themselves, as many-body currents and single-nucleon electromagnetic form factors need to be accounted for. 

Neglecting small effects, the charge radius $\left\langle r_{\rm ch}^2\right\rangle$ can be expressed in terms of the point proton radius $\left\langle r_{\rm pt}^2\right\rangle$~\cite{friar:1997}
\begin{align}
	\left\langle r_{\rm ch}^2\right\rangle=\left\langle r_{\rm pt}^2\right\rangle+\left\langle R_p^2\right\rangle+\frac{N}{Z}\left\langle R_n^2\right\rangle+\frac{3\hbar^2}{4M_p^2 c^2},
	\label{eq:rch}
\end{align}
where $\left\langle R_p^2\right\rangle=0.770(9)\,\rm fm^2$ is the proton radius~\cite{beringer:2012}, $\left\langle R_n^2\right\rangle=-0.116(2)\,\rm fm^2$ is the neutron radius~\cite{beringer:2012}, and $3\hbar^2/(4M_p^2 c^2)\simeq 0.033\,\rm fm^2$ is the Darwin-Foldy term. Charge radii in $^4$He, $^{16}$O, and $^{40}$Ca for both AV18 and AV18+UIX are reported in \cref{tab:he4,tab:o16,tab:ca40}, respectively. In $^4$He AV18 produces a charge radius larger than the experimental value. However, the $3N$ force shrinks the nucleus, improving the agreement with experiment. Both CVMC values are reasonably consistent with the VMC ones. In $^{16}$O and $^{40}$Ca instead, the $N\!N$ interaction alone results in too small radii, while the UIX potential increases them towards and above their experimental values. This is consistent with the observation that, as opposed to $^4$He, for $A=16$ and $A=40$ the net effect of the UIX potential is to make the systems more loosely bound. 

\begin{table}[b]
\centering
\caption[]{Total energies (in MeV/$A$) and charge radii (in fm) in $^{16}$O for different potentials.}
\label{tab:o16}
\begin{tabular}{l l c c}
\hline\hline
obs & potential & CVMC & exp \\
\hline
\multirow{2}{*}{$\left\langle E\right\rangle$} & AV18 & $-5.51(2)$ & \multirow{2}{*}{$-7.98$} \\
& AV18+UIX & $-5.15(2)$ & \\ [0.2cm] 
\multirow{2}{*}{$\sqrt{\left\langle r_{\rm ch}^2\right\rangle}$} & AV18 & $2.538(2)$ & \multirow{2}{*}{$2.699(5)$} \\
& AV18+UIX & $2.745(2)$ & \\
\hline\hline
\end{tabular}
\vspace{0.5cm}
\centering
\caption[]{Total energies (in MeV/$A$) and charge radii (in fm) in $^{40}$Ca for different potentials.}
\label{tab:ca40}
\begin{tabular}{l l c c}
\hline\hline
obs & potential & CVMC & exp \\
\hline
\multirow{2}{*}{$\left\langle E\right\rangle$} & AV18 & $-5.88(10)$ & \multirow{2}{*}{$-8.55$} \\
& AV18+UIX & $-4.92(10)$ & \\ [0.2cm] 
\multirow{2}{*}{$\sqrt{\left\langle r_{\rm ch}^2\right\rangle}$} & AV18 & $3.361(2)$ & \multirow{2}{*}{$3.478(1)$} \\
& AV18+UIX & $3.617(2)$ & \\
\hline\hline
\end{tabular}
\end{table}

The single-nucleon, two-nucleon, and two-nucleon operator point densities are defined as
\begin{align}
	\!\!\rho_N(r)&=\frac{1}{4\pi r^2}\frac{\langle\Psi_V|\sum_i\delta(\tilde{r}_i)\,\mathcal P_{N_i}\,|\Psi_V\rangle}{\langle\Psi_V|\Psi_V\rangle} , \label{eq:rho_N} \\[0.1cm]
    \!\!\rho_{\rm{NN}}(r)&=\frac{1}{4\pi r^2}\frac{\langle\Psi_V|\sum_{i<j}\delta(r-r_{ij})\,\mathcal P_{N_i}\mathcal P_{N_j}\,|\Psi_V\rangle}{\langle\Psi_V|\Psi_V\rangle},\!  \label{eq:rho_NN}\\[0.1cm]
    \!\!\rho_{2,p}(r)&=\frac{1}{4\pi r^2}\frac{\langle\Psi_V|\sum_{i<j}\delta(r-r_{ij})\,\mathcal O_{ij}^p\,|\Psi_V\rangle}{\langle\Psi_V|\Psi_V\rangle} ,
\end{align}
where $N=p,n$, $\mathcal P_{N_i}=(1\pm\tau_{z_i})/2$ are isospin projection operators, and the operators $\mathcal O_{ij}^p$ are given in \cref{eq:v6}. With these definitions, $\rho_N(r)$ is normalized to the number of protons or neutrons, and $\rho_{\rm{NN}}$ to the number of $pp$, $np$ or $nn$ pairs. Note that for some of the alternative expansion schemes normalization is ensured order-by-order by construction (either defining a ``number conserving'' expansion~\cite{alvioli:2005,alvioli:2008,alvioli:2013,alvioli:2016}, or requiring a normalization factor~\cite{ryckebusch:2015}). In our expansion scheme, the central one- and two-body densities are properly normalized order-by-order. The first term of the corresponding cluster expansion carries the full normalization, and higher order contributions integrate to zero within Monte Carlo statistical errors. This reflects the fact that at every order a $3A$ dimensional integral is performed. The normalization of the two-body operator densities is instead recovered only at convergence, and each order of the cluster expansion contributes to it.

\begin{table}[b]
\centering
\caption[]{Total energies (in MeV/$A$) and charge radii (in fm) in $^4$He and $^{16}$O for the AV6$^\prime$ potential. The electromagnetic term $v_{ij}^\gamma$ is not included. AFDMC energies are taken from Ref.~\cite{gandolfi:2014_prc}.}
\label{tab:av6p}
\begin{tabular}{l c c c}
\hline\hline
obs & nucleus & CVMC & AFDMC \\
\hline
\multirow{2}{*}{$\left\langle E\right\rangle$}	& $^4$He & $-6.53(1)$ & $-6.77(1)$ \\
												& $^{16}$O & $-6.79(3)$ & $-7.23(2)$ \\ [0.2cm]
\multirow{2}{*}{$\sqrt{\left\langle r_{\rm ch}^2\right\rangle}$}	& $^4$He & $1.678(3)$ & $1.674(9)$ \\
																	& $^{16}$O & $2.580(2)$ & $2.52(3)$ \\
\hline\hline
\end{tabular}
\end{table}

\begin{figure}[t]
	\centering
	\includegraphics[width=\linewidth]{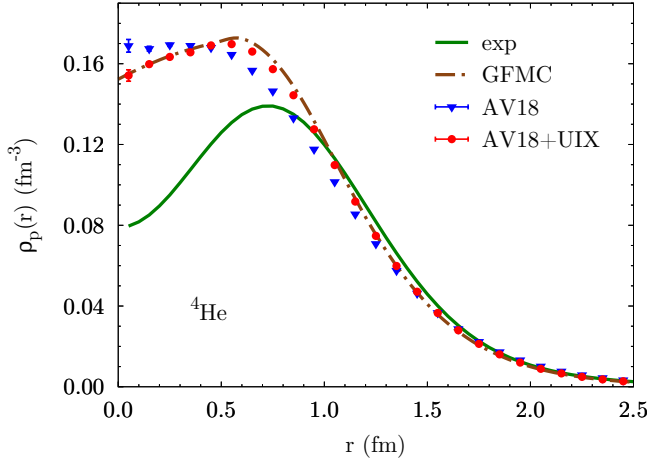}
	\caption[]{Point proton densities in $^4$He. The solid green line refers to the ``experimental'' result; see text for details. The dash-dotted brown line is the GFMC result for AV18+UIX~\cite{wiringa:rhor}.}
	\label{fig:rhor_he4}
\end{figure}

\begin{figure}[b]
	\centering
	\includegraphics[width=\linewidth]{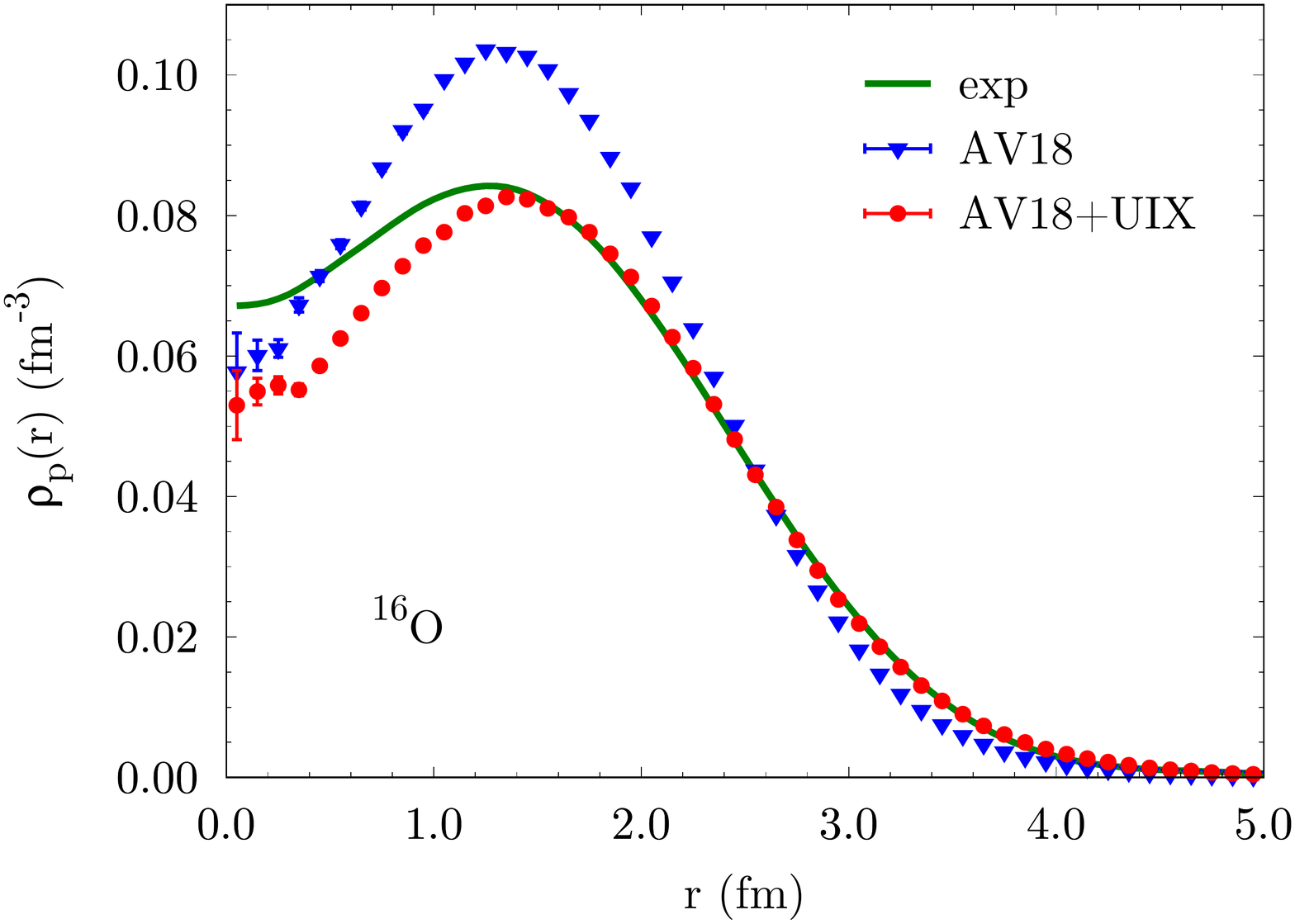}
	\caption[]{Point proton densities in $^{16}$O. The green line refers to the ``experimental'' result; see text for details.}
	\label{fig:rhor_o16}
\end{figure}

\begin{figure}[htb]
	\centering
	\includegraphics[width=\linewidth]{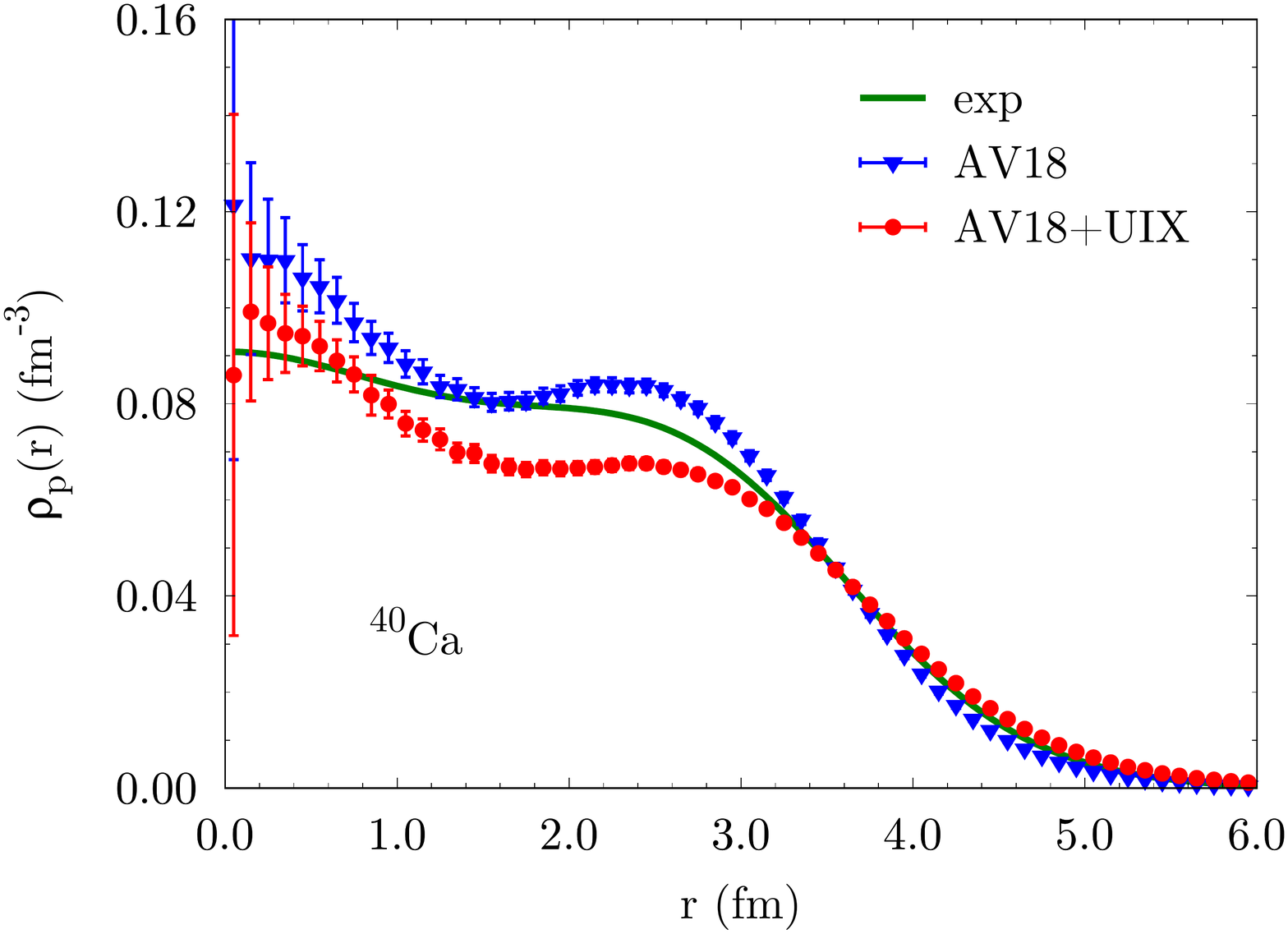}
	\caption[]{Point proton densities in $^{40}$Ca. The green line refers to the ``experimental'' result; see text for details.}
	\label{fig:rhor_ca40}
\end{figure}

\Cref{fig:rhor_he4,fig:rhor_o16,fig:rhor_ca40} show the point proton densities of $^4$He, $^{16}$O, and $^{40}$Ca, respectively, obtained with the AV18 and AV18+UIX interactions. They are compared to the values obtained from the ``Sum-of-Gaussians'' parametrization of the charge densities given in Ref.~\cite{devries:1987} by unfolding the nucleon form factors and subtracting the small contribution of the neutrons. As discussed at length in Sec.~\ref{subsec:ff_sr}, neglecting two-body meson exchange currents (MECs) is likely to have little effect in $^{16}$O and $^{40}$Ca. On the other hand, MECs are important in the description of the $^4$He elastic form factor, from which the charge densities are extracted. Hence, the discrepancy between theory and experiment of \cref{fig:rhor_he4} does not have to be ascribed to deficiencies of the CVMC wave function. In fact, the $^4$He point proton density obtained within CVMC for AV18+UIX agrees very well with the GFMC result for the same interaction.

For the lightest system the effect of the $3N$ force on the density is not dramatic, as expected by looking at the small difference in the charge radii of \cref{tab:he4}. In oxygen and calcium, instead, the addition of the UIX potential pushes the nucleons far away from the center of mass. For both systems the density at small distances is substantially depleted, with a $\simeq 25\%$ reduction of both the peak in $^{16}$O at $1.4\,\rm fm$ and the plateau in $^{40}$Ca around $2\,\rm fm$. Remarkably, this effect results in a better description of the structure of $^{16}$O, for which the AV18+UIX prediction of the charge radius is less than $2\%$ different from the experimental value, as shown in \cref{tab:o16}. However, the situation is different in the case of $^{40}$Ca, for which the employed $3N$ force is too repulsive and pushes the nucleons towards the surface of the nucleus yielding an excessively large charge radius, as in \cref{tab:ca40}. The point proton density of $^{40}$Ca turns out to be $0.8\,\rm fm ^{-3}$ at $1\,\rm fm$, and $0.7\,\rm fm ^{-3}$ in the plateau after $1.6\,\rm fm$. These values are consistent with the saturation density of SNM obtained employing the same Hamiltonian, while AV18 alone significantly overpredicts the saturation density~\cite{akmal:1998}.

\begin{figure}[b]
	\centering
	\includegraphics[width=\linewidth]{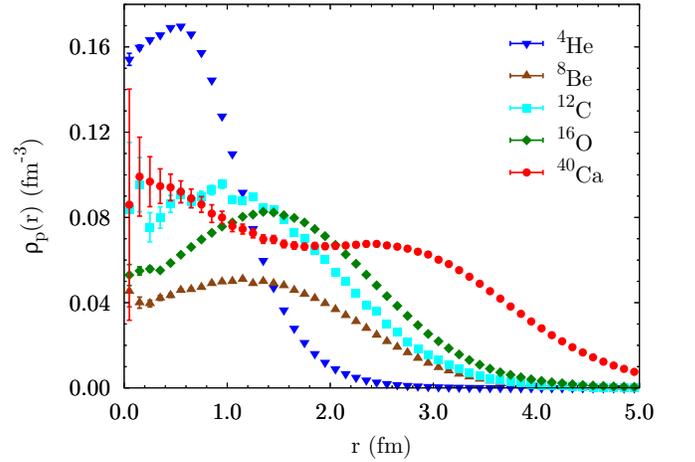}
	\caption[]{Point proton densities for AV18+UIX. $^4$He, $^{16}$O, and $^{40}$Ca are the results of this work. $^8$Be and $^{12}$C are VMC results collected in~\cite{wiringa:rhor}.}
	\label{fig:rhor_A_av18+uix}
\end{figure}

\begin{figure}[t]
	\centering
	\includegraphics[width=\linewidth]{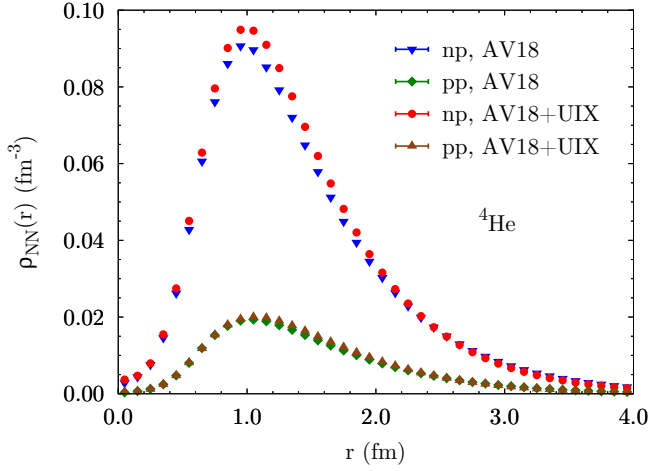}
	\caption[]{Two-nucleon densities of $^4$He.}
	\label{fig:rhor12_he4}
\end{figure}

\begin{figure}[b]
	\centering
	\includegraphics[width=\linewidth]{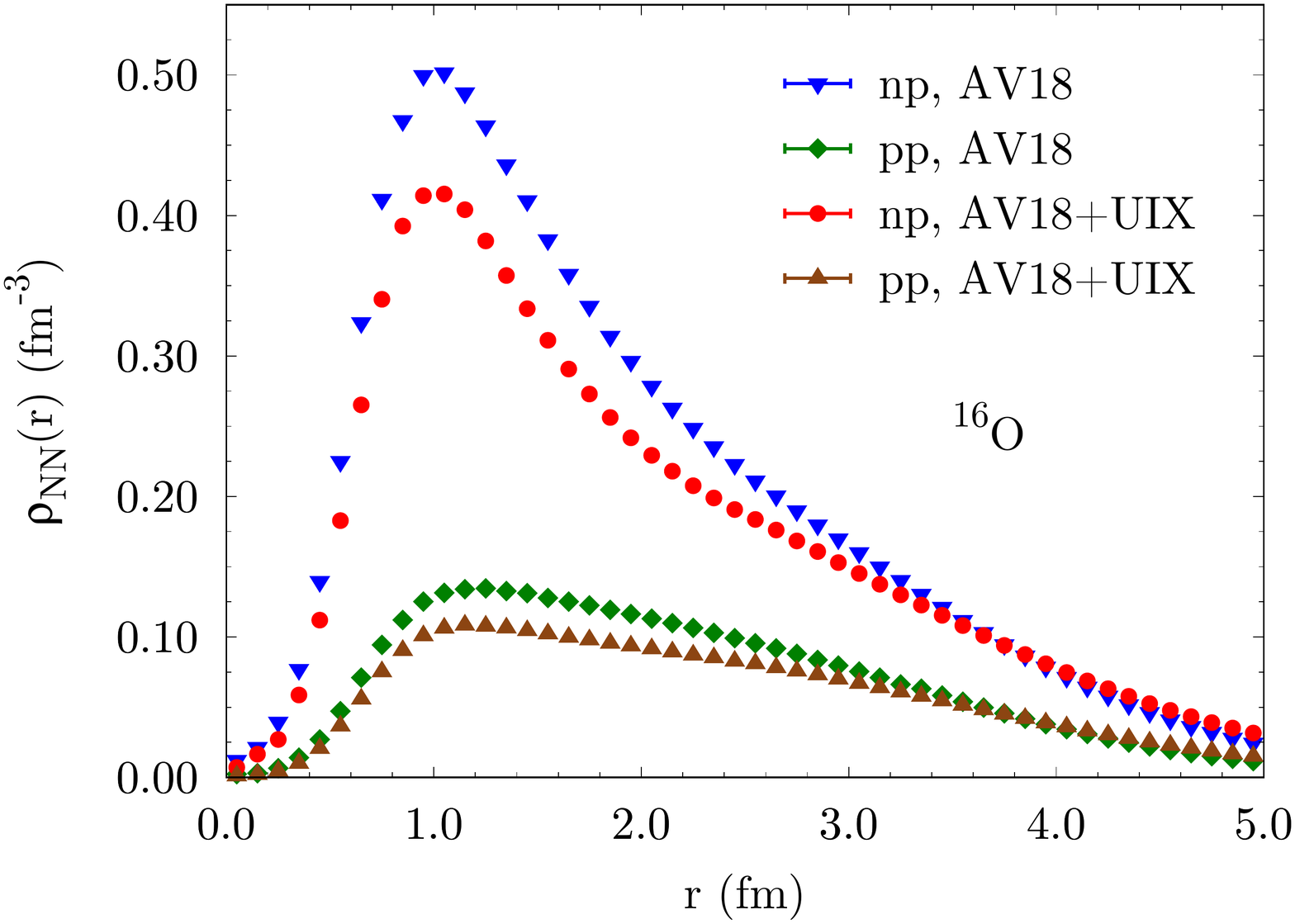}
	\caption[]{Two-nucleon densities of $^{16}$O.}
	\label{fig:rhor12_o16}
\end{figure}

It is interesting to compare the densities of these nuclei in which $\alpha$-clustering can potentially occur. In \cref{fig:rhor_A_av18+uix} we collect the CVMC results for the point proton densities of $^4$He, $^{16}$O, and $^{40}$Ca obtained with AV18+UIX together with those for $^8$Be and $^{12}$C coming from VMC calculations using the same interaction~\cite{wiringa:rhor}. The $^4$He density shows a very large point density at small distance. When integrated over the volume, about half the nucleons reside inside $1.25\,\rm fm$, where the density is above $0.08\,\rm fm^{-3}$. The $^8$Be density has a low, broad peak with half the nucleons residing inside $2.25\,\rm fm$, consistent with a two-$\alpha$ cluster structure as observed in Fig.~15 of Ref.~\cite{wiringa:2000}. The $^{12}$C density peaks at a slightly smaller distance and noticeably higher value, with a larger dip at the center. This is consistent with a more tightly bound three-$\alpha$ cluster---either in a triangular configuration with a low-density region at the center of mass, or alternatively with one $\alpha$ in the $s$-shell and two $\alpha$'s in the $p$-shell.  Similarly, $^{16}$O can be viewed as a tetrahedral four-$\alpha$ cluster with the $\alpha$'s at somewhat greater distance from the center of mass, or as one $s$-shell and three $p$-shell $\alpha$'s with a larger dip-peak difference than in $^{12}$C.  The $^{40}$Ca density is more complicated, but might be thought of as two $s$-shell $\alpha$'s giving a larger central peak, while three $p$-shell and five $d$-shell $\alpha$'s give a broad shoulder at $1-3\,\rm fm$.

\begin{figure}[t]
	\centering
	\includegraphics[width=\linewidth]{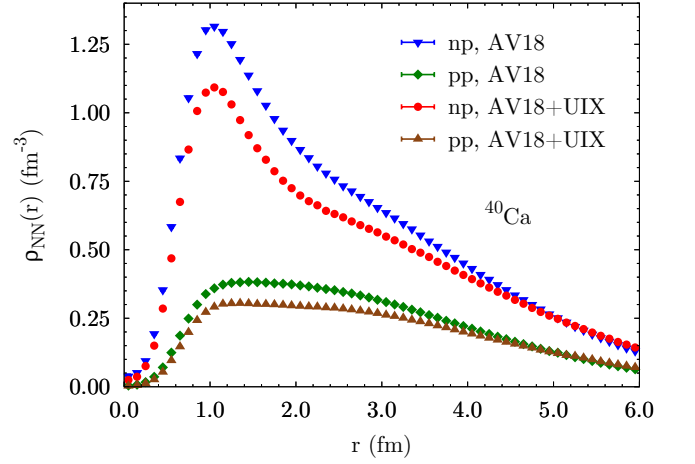}
	\caption[]{Two-nucleon densities of $^{40}$Ca.}
	\label{fig:rhor12_ca40}
\end{figure}

\begin{figure}[b]
	\centering
	\includegraphics[width=\linewidth]{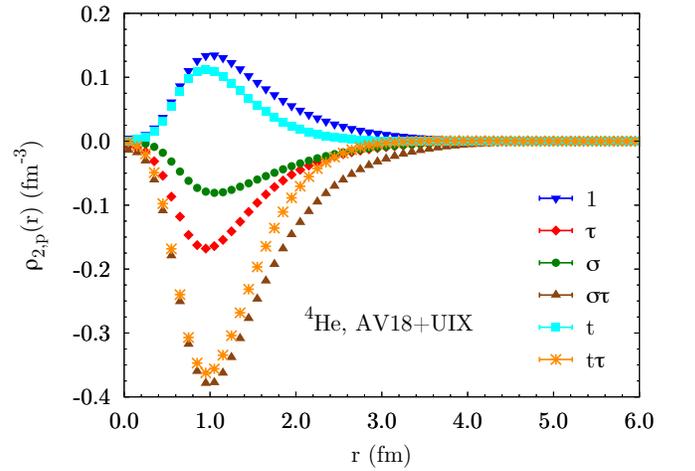}
	\caption[]{Operator two-nucleon densities in $^4$He. $\mathbbm{1},\,\tau,\,\sigma,\,\sigma\tau,\,t,\,t\tau$ correspond to operators $p=1,\ldots,6$ in Eq.~(\ref{eq:v6}).}
	\label{fig:rhor12op_he4_av18+uix}
\end{figure}

The two-nucleon point densities of $^4$He, $^{16}$O, and $^{40}$Ca are reported in \cref{fig:rhor12_he4,fig:rhor12_o16,fig:rhor12_ca40}, respectively, for both AV18 and AV18+UIX. Upper and lower curves refer to $np$ and $pp$ pairs, respectively. The fact that $\rho_{\rm{NN}}$ is very small for $r\simeq 0$ is a consequence of the repulsive core of the $N\!N$ potential. As observed for the point-proton densities, the effect of the $3N$ force on the two-nucleon densities is appreciably different in light- and medium-heavy systems. In $^4$He the $pp$ density is almost unchanged, while the $np$ density is enhanced around the peak at $1.1\,\rm fm$. In heavier systems there is a severe depletion of both $pp$ and $np$ densities, again due to the peculiar repulsive effect of the UIX potential that tends to push nucleons apart. 

Both figures and tables for the CVMC single-nucleon and two-nucleon point densities for $^{16}$O and $^{40}$Ca, together with the VMC results for $A\leq12$, are available online~\cite{wiringa:rhor,wiringa:rhor12}.

\begin{figure}[t]
	\centering
	\includegraphics[width=\linewidth]{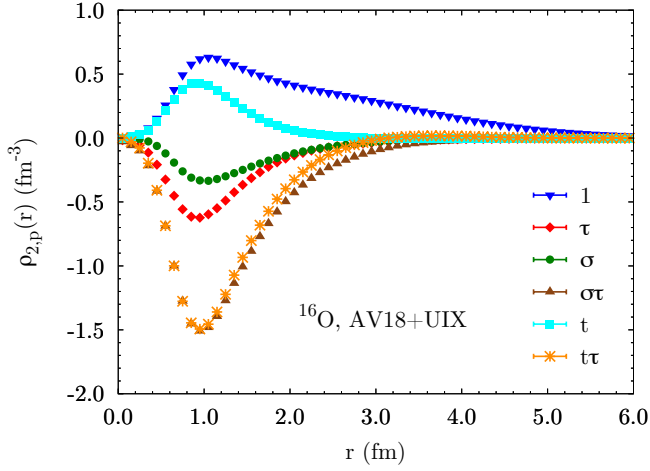}
	\caption[]{Operator two-nucleon densities in $^{16}$O.}
	\label{fig:rhor12op_o16_av18+uix}
\end{figure}

\begin{figure}[b]
	\centering
	\includegraphics[width=\linewidth]{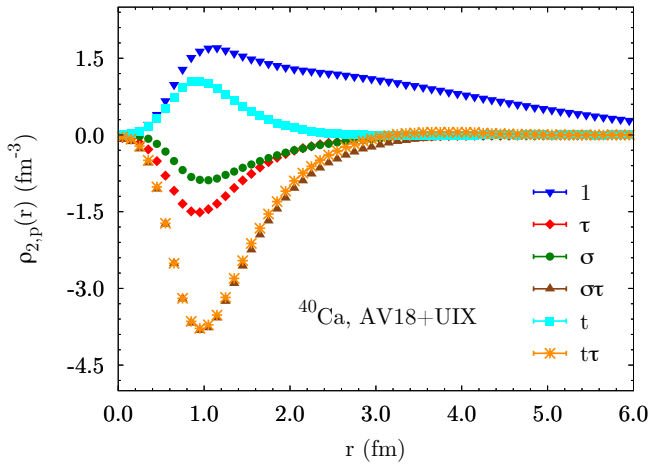}
	\caption[]{Operator two-nucleon densities in $^{40}$Ca.}
	\label{fig:rhor12op_ca40_av18+uix}
\end{figure}

The two-nucleon operator point densities are shown in \cref{fig:rhor12op_he4_av18+uix,fig:rhor12op_o16_av18+uix,fig:rhor12op_ca40_av18+uix} for $^4$He, $^{16}$O, and $^{40}$Ca, respectively. It can be observed that the larger the system, the wider the range of central two-body density. In fact, the central two-body operator density is just the sum of $pp$, $np$, and $nn$ densities in \cref{fig:rhor12_he4,fig:rhor12_o16,fig:rhor12_ca40}. On the other hand, spin-isospin densities are appreciably nonvanishing only for $r \lesssim 3.5\,\rm fm$, and are largely independent of the nucleus, with the position of the peaks situated around $1\,\rm{fm}$. This extends the results of Ref.~\cite{feldmeier:2011}, where the two-body densities normalized at short distances in $A=3$ and $A=4$ systems exhibit a universal behavior up to about $1\,\rm{fm}$ in all nuclei. Among the spin-isospin densities, $\rho_{2,\sigma\tau}$ and $\rho_{2,t\tau}$ are characterized by longer ranges and larger amplitudes, as they arise from the one-pion-exchange part of the $N\!N$ interaction. These results are qualitatively consistent with the findings of Ref.~\cite{alvioli:2005}, although no $3N$ forces were employed in that work. The peak values of the $\rho_{2,\sigma\tau}$ and $\rho_{2,t\tau}$ scale as $1:4:10$ for ${^4\rm He}:{^{16}\rm O}:{^{40}\rm Ca}$, or just as the number of $\alpha$-particle clusters.

\subsection{Momentum distributions}

The probability of finding a proton or neutron with momentum $\bm k$ is proportional to the momentum distribution,
\begin{align}
	n_N(\bm k)&=\int d\bm r_1^\prime \,d\bm r_1^{\phantom{\prime}}\,d\bm r_2^{\phantom{\prime}} \cdots d\bm r_A^{\phantom{\prime}}\,\Psi^\dagger(\bm r_1^\prime,\bm r_2^{\phantom{\prime}},\ldots,\bm r_A^{\phantom{\prime}} ) \nonumber \\
    &\times e^{-i\bm k\cdot(\bm r_1^{\phantom{\prime}}-\bm r_1^\prime)}\,\mathcal P_{N_1}\,\Psi(\bm r_1^{\phantom{\prime}},\bm r_2^{\phantom{\prime}},\ldots, \bm r_A^{\phantom{\prime}} ) , \label{eq:nofk_1}
\end{align}
which is normalized as
\begin{align}
	\mathcal N_N=\int \frac{d\bm k}{(2\pi)^3}\,n_N(\bm k) ,
\end{align}
$\mathcal N_N$ being the number of protons or neutrons ($\mathcal N_p\equiv Z$). In this work we present results for symmetric nuclei implying $n_p(\bm k)=n_n(\bm k)$. Equation~(\ref{eq:nofk_1}) can be rewritten as
\begin{align}
    n_N(\bm k)&=\frac{1}{A}\sum_i\int d\bm r_1\cdots d\bm r_i\cdots d\bm r_A\int d\Omega_x \int_0^{x_{\max}} dx\,x^2 \nonumber \\ 
    & \times \Psi^\dagger(\bm r_1,\ldots,\bm r_i,\ldots,\bm r_A)\,e^{-i\bm k\cdot\bm x} \nonumber \\
    & \times \mathcal P_{N_i}\,\Psi(\bm r_1,\ldots,\bm r_i+\bm x,\ldots,\bm r_A) \,. \label{eq:nofk_2}
\end{align}

The Fourier transform can be computed by Monte Carlo integration. Spatial configurations are sampled as explained in Sec.~\ref{sec:sampl}. The average over all particles $i$ in each configuration is then performed, and for each particle, a grid of Gauss-Legendre points $x_i$ is used to compute the Fourier transform. The polar angle $d\Omega_x$ is also sampled by Monte Carlo integration, with a randomly chosen direction for each particle in each configuration. For all the nuclei under study we calculated $n(\bm k)$ up to $k=10\,\rm fm^{-1}$, integrating to $x_{\max}=20\,\rm fm$ using 200 Gauss-Legendre points.

\begin{figure}[t]
	\centering
	\includegraphics[width=\linewidth]{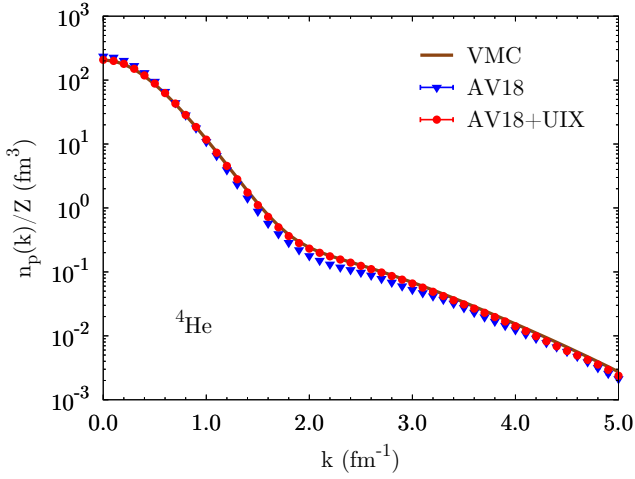}
	\caption[]{Proton momentum distributions in $^4$He. Averages are calculated on $\mathcal N_c=10^7$ configurations. The brown line is the VMC result for AV18+UIX~\cite{wiringa:rhok}.} 
	\label{fig:nofk_he4}
\end{figure}

\begin{figure}[b]
	\centering
	\includegraphics[width=\linewidth]{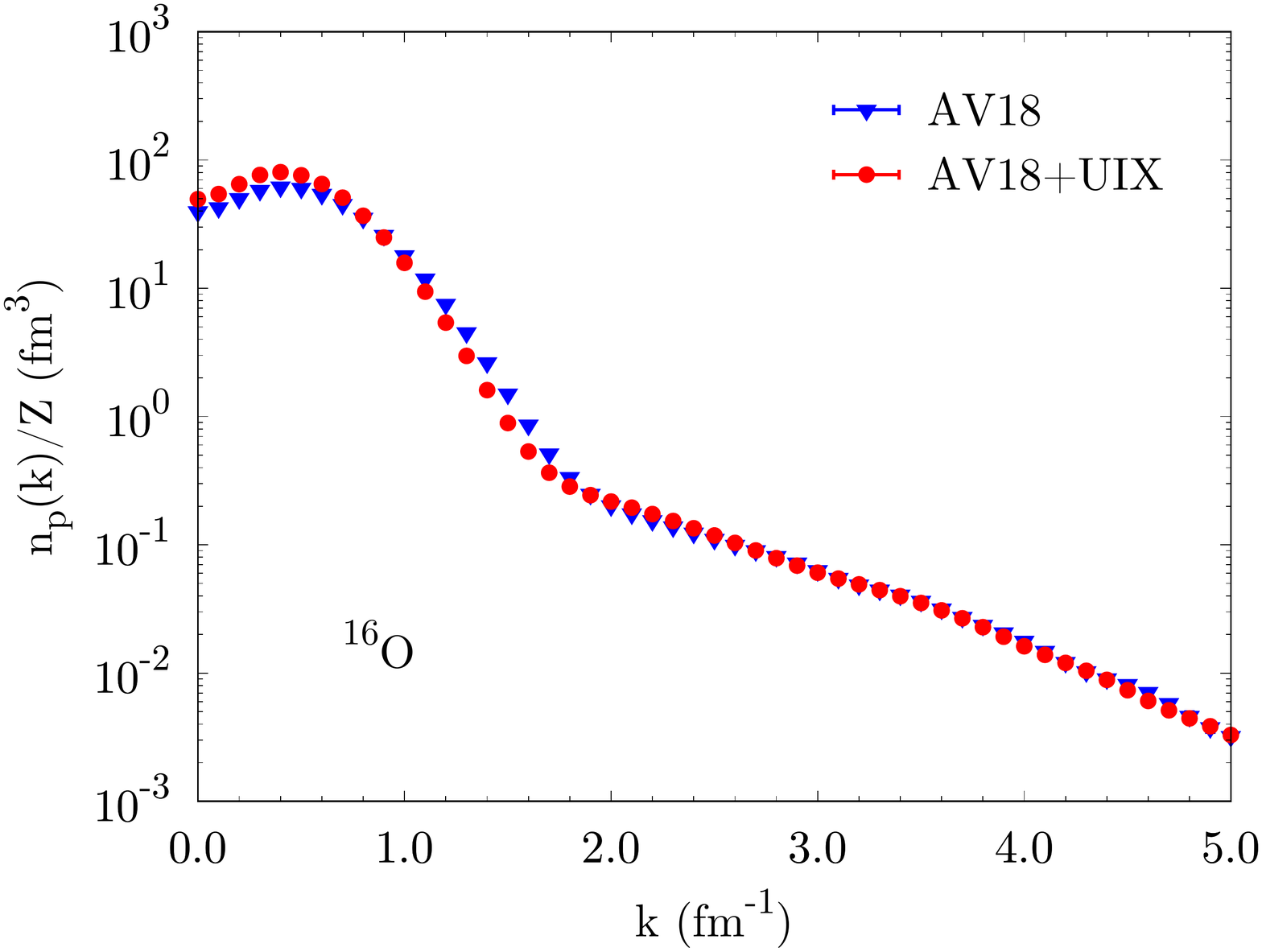}
	\caption[]{Proton momentum distributions in $^{16}$O. Averages are calculated on $\mathcal N_c=10^7$ configurations for AV18, and on $\mathcal N_c=8\times10^6$ configurations for AV18+UIX.}
	\label{fig:nofk_o16}
\end{figure}

\begin{figure}[t]
	\centering
	\includegraphics[width=\linewidth]{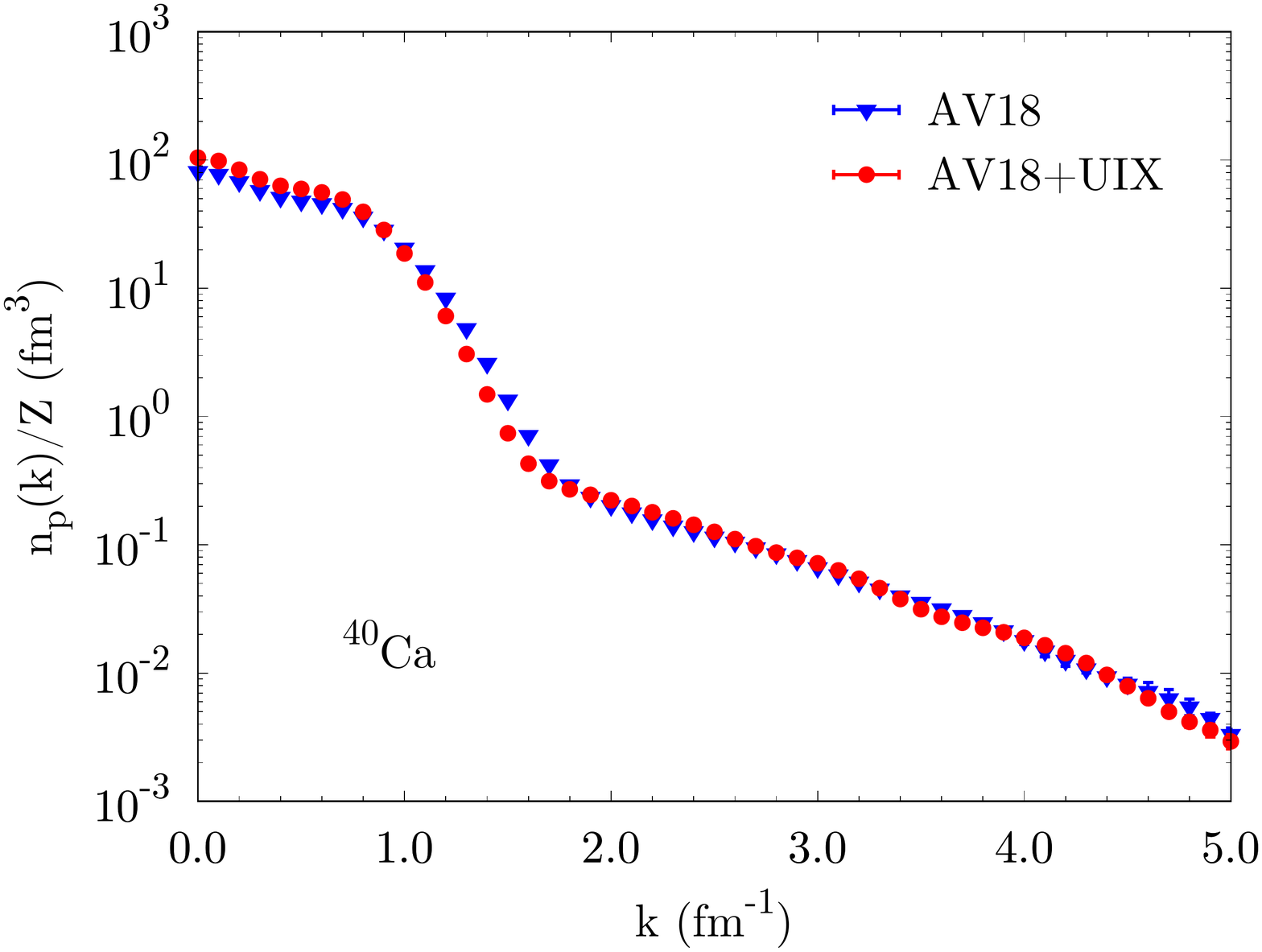}
	\caption[]{Proton momentum distributions in $^{40}$Ca. Averages are calculated on $\mathcal N_c=10^7$ configurations for AV18, and on $\mathcal N_c=8\times10^6$ configurations for AV18+UIX.}
	\label{fig:nofk_ca40}
\end{figure}

\begin{figure}[b]
	\centering
	\includegraphics[width=\linewidth]{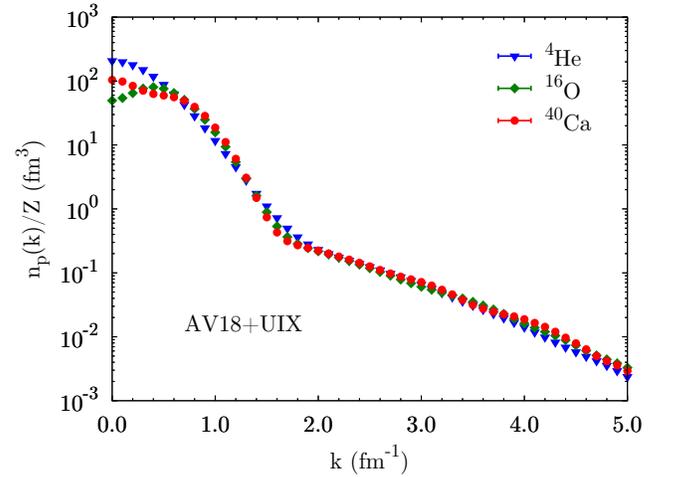}
	\caption[]{Proton momentum distributions for AV18+UIX.}
	\label{fig:nofk_A_av18+uix}
\end{figure}

As reported in Ref.~\cite{pieper:1992}, with the employed expansion the three-body clusters give small contribution to the momentum distribution. In this work, $n_N(\bm k)$ is evaluated up to three-body cluster and then extrapolated using \cref{eq:Cext2}. In order to save computing time, spin-orbit correlations are turned off in the calculation of the momentum distribution. This approximation, also used in standard VMC calculations~\cite{wiringa:2014}, is justified by the small effect of spin-orbit correlations on $n_N(\bm k)$ compared to the first six operators of the two-body correlations. The results for $^4$He, $^{16}$O, and $^{40}$Ca are shown in \cref{fig:nofk_he4,fig:nofk_o16,fig:nofk_ca40}, respectively. For $A=4$ the VMC result for AV18+UIX~\cite{wiringa:rhok} is also displayed for comparison. The proton momentum distributions are reported for both AV18 and AV18+UIX. The $3N$ force makes only small changes to $n_N(\bm k)$. Near $k=2\,\rm fm^{-1}$ the momentum distribution manifests a sharp change in slope, as previously observed in both light-~\cite{wiringa:2014} and medium-mass~\cite{alvioli:2008} nuclei. This is attributed to the strong tensor correlations induced by the one-pion-exchange part of the $N\!N$ potential, further enhanced by the two-pion-exchange part of the $3N$ potential, when included. At higher momentum, the tail of $n_N(\bm k)$ manifests the expected universal behavior determined by the short-range correlations, i.e., by the short-range structure of the employed Hamiltonian, as shown in \cref{fig:nofk_A_av18+uix} and discussed at length in a number of other works~\cite{feldmeier:2011,alvioli:2012,alvioli:2013,alvioli:2016}. Such universality refers to the independence of the tail with respect to the specific nucleus. On the other hand, the high-momentum tail strongly depends on the nuclear interaction model. The recently developed local chiral interactions, which are significantly softer than the phenomenological interactions employed in this work, yield a momentum distribution characterized by weaker high-momentum components~\cite{gandolfi:2017} than those of \cref{fig:nofk_he4,fig:nofk_o16,fig:nofk_ca40}.

Compared to the other local or nearly local operators, like the kinetic energy, the potential energy, and the densities, the momentum distribution is strictly a nonlocal operator. In order to check the convergence of the cluster expansion for such operator we computed the kinetic energy by integrating the momentum distribution
\begin{align}
	E_N^{\rm kin}(k)=\frac{\hbar^2}{2m_N} 4\pi \int_0^k dk^\prime\,k^{\prime\,4}\,n_N(k^\prime) ,
\end{align}
for each order of the expansion. The contributions for $^{16}$O with AV18+UIX up to $k=10\,\rm fm^{-1}$ are $16.3(2)\,\rm MeV/A$ for one-body cluster, $16.0(5)\,\rm MeV/A$ for two-body cluster, and $-4.4(4)\,\rm MeV/A$ for three-body cluster. The integration of the extrapolated $n_N(\bm k)$ leads to $28.9(6)\,\rm MeV/A$. This is compatible with the cluster contributions reported in the first line of \cref{tab:o16_av18+uix}. The missing 4- to 16-body cluster contributions to the integrated kinetic energy, that account for $\simeq1\,\rm MeV/A$, are fully recovered by the extrapolation of $n_N(\bm k)$. This validates the convergence of the expansion and confirms the negligible effect of spin-orbit correlations on the momentum distribution. Similar outcomes are found for the other nuclei considered in this work. The errors on the integrated kinetic energies are larger than those of the direct calculation because of the propagation of uncertainties in the integration of $n_N(\bm k)$, which above $5\,\rm fm^{-1}$ has large statistical errors due to the cancellation of positive and negative small cluster contributions. However, as discussed in the next paragraph, the integrated strength of the momentum distribution saturates before $5\,\rm fm^{-1}$. Simulations for $n_N(\bm k)$ have been thus carried out with good statistics up to that momentum value, using up to $10^7$ Monte Carlo configurations.

\begin{figure}[t]
	\centering
	\includegraphics[width=\linewidth]{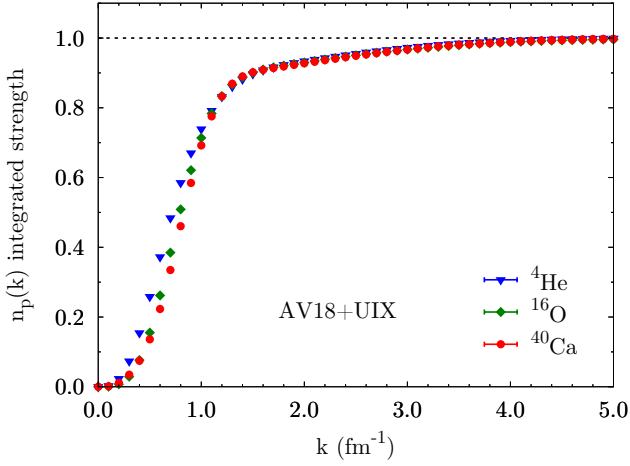}
	\caption[]{Integrated strengths for AV18+UIX. See text for details.}
	\label{fig:nofk_strng_av18+uix}
\end{figure}

The momentum distribution integrated strength as a function of $k$ is reported in \cref{fig:nofk_strng_av18+uix} for AV18+UIX. At low momentum it decreases as $A$ increases, because the nuclei become more tightly bound, and the fraction of nucleons at low momentum decreases. At $k=2\,\rm fm^{-1}$ for all the systems analyzed, the integrated strength is already $\simeq93\%$ of the total, and it becomes $\simeq99\%$ at $k=4\,\rm fm^{-1}$. Less than $1\%$ of the total strength is given by the tail of the momentum distribution above $4\,\rm fm^{-1}$.

The figures and the tables for the CVMC momentum distributions for $^{16}$O and $^{40}$Ca, together with the VMC results for $A\leq12$, are available online~\cite{wiringa:rhok}.

\subsection{Charge form factors and Coulomb sum rules}
\label{subsec:ff_sr}
The double differential cross section of the inclusive electron-nucleus scattering process in which an electron of initial four-momentum $k_e=(\bm{k}_e,E_e)$ scatters off a nuclear target to a state of four-momentum $k_{e}^\prime=(\bm{k}_{e}^\prime,E_{e}^\prime)$, the hadronic final state being undetected, can be written in the one-photon-exchange approximation as
\begin{align}
	\!\!\!\frac{d^2\sigma}{d E_{e^\prime} d\Omega_{e^\prime}}\! =\!\left( \frac{d \sigma}{d\Omega_{e^\prime}} \right)_{\!\rm M}\!\Big[  A_L  R_L({\bm q},\omega) 
	 + A_T  R_T({\bm q},\omega) \Big] ,
	\label{eq:inclusive}
\end{align}
where 
\begin{align}
	A_L = \left( \frac{Q^2}{\bm q^2}\right)^2 , \qquad A_T = \frac{1}{2}\frac{Q^2}{\bm q^2}+\tan^2\frac{\theta_{e^\prime}}{2} , 
\end{align}
and
\begin{align}
	\left(\frac{d \sigma}{d \Omega_{e^\prime}}\right)_{\!\rm M}=\left[\frac{\alpha\cos{\displaystyle\frac{\theta_{e^\prime}}{2}}}{2 E_e\sin^2{\displaystyle\frac{\theta_{e^\prime}}{2}}}\right]^2
    \label{Mott}
\end{align}
is the Mott cross section. In the above expressions, $\alpha\simeq1/137$ is the fine structure constant, $d\Omega_{e^\prime}$ is the differential solid angle in the direction of $\bm k_{e^\prime}$, $q=k_e-k_{e^\prime}=(\bm q,\omega)$ is the four-momentum transfer, and $Q^2=-q^2=\bm q^2-\omega^2$. The longitudinal and transverse response functions are defined as
\begin{align}
	R_\alpha (\bm{q},\omega)&=\sum_f \langle f | j_\alpha({\bm q},\omega) |0\rangle 
	\langle f | j_\alpha({\bm q},\omega) |0\rangle^* \nonumber\\
	& \times \delta(E_f-\omega-E_0),\qquad \alpha=L,T,
	\label{eq:res_def}
\end{align}
where $|0\rangle$ and $|f\rangle$ represent the nuclear initial and final
states of energies $E_0$ and $E_f$, and $j_L({\bm q},\omega)$ and $j_T({\bm q},\omega)$ are the electromagnetic charge and current operators, respectively. 

Recently, the quasielastic electromagnetic response functions of $^4$He and $^{12}$C have been computed within GFMC using realistic nuclear two- and three-body forces and consistent one- and two-body electroweak currents~\cite{lovato:2015,lovato:2016}. Besides the transverse enhancement brought about by two-body current contributions, the authors of Ref.~\cite{lovato:2016} have found no evidence of in-medium modification of the nucleon form factor in the analysis of the longitudinal response function of $^{12}$C. This is at variance with the findings of Ref.~\cite{cloet:2016}, where changes to the proton Dirac form factor induced by the nuclear medium leads to a dramatic quenching of the Coulomb sum rule,
\begin{align}
	S_L(\bm{q})=\frac{1}{Z}\int_{w_{\rm th}^+}^\infty d\omega\,\frac{R_L(\bm{q},\omega)}{G_E^{p\,2}(Q^2)} ,
	\label{eq:SL_def}
\end{align}
where $\omega_{\rm th}$ is the energy transfer corresponding to the inelastic threshold, and $G_E^p(Q^2)$ is the proton electric form factor evaluated at four-momentum transfer $Q^2$. 

The one-body charge operator employed in the GFMC calculations has the standard expressions obtained from a relativistic reduction of the time component of the covariant single-nucleon current,
\begin{align}
	j_L({\bm q},\omega)=& \Bigg[ \frac{\epsilon_i(Q^2)}{\sqrt{1+Q^2/(4 m^2)}}\nonumber\\
	&-i\,\frac{2\mu_i(Q^2)-\epsilon_i(Q^2)}{4 m^2}\,\bm q\cdot ({\bm \sigma}_i \times {\bm p}_i)\Bigg]e^{i\bm q\cdot\bm r_i},
	\label{eq:jL_def}
\end{align}
with
\begin{align}
	\begin{aligned}
		\epsilon_i(Q^2)&= G^p_{E}(Q^2)\frac{1+	\tau_{z_i}}{2} + G_E^n(Q^2)\frac{1-\tau_{z_i}}{2}, \\
		\mu_i(Q^2) &= G_M^p(Q^2)\frac{1+\tau_{z_i}}{2}+ G_M^n(Q^2)\frac{1-\tau_{z_i}}{2}\,.
	\label{form:fact}
	\end{aligned}
\end{align}
In this work we adopted Kelly's parametrization~\cite{kelly:2004} for the nucleon electric and magnetic form factors $G^{p,n}_{E,M}$.

In $R_L(\bm{q},\omega)$, the $\omega$ dependence enters via the energy-conserving $\delta$ function and the four-momentum transfer $Q^2$ of the electroweak form factors of the nucleon. The latter can be removed by evaluating these form factors at $Q^2_{\rm qe} = \bm{q}^2 - \omega^2_{\rm qe}$, where $\omega_{\rm qe}$ is the energy transfer corresponding to the quasielastic peak, and by dividing the response by the factor $G_E^{p\,2}(Q_{\rm qe}^2)$. Therefore, the Coulomb sum rule can be very well approximated by the following ground-state expectation value
\begin{align}
	S_L(\bm q)=&\frac{1}{Z} \Big[ \langle 0|\mathcal O_L^\dagger({\bm q}) \,\mathcal O_L({\bm q}) |0\rangle\nonumber\\
	&-|\langle 0;{\bm q}|\mathcal O_L({\bm q}) |0 \rangle|^2 \Big]  ,
	\label{eq:SL_gs}
\end{align}
where $\mathcal O_L({\bm q})=j_L({\bm q},\omega_{\rm qe})/G_E^p(Q_{\rm qe}^2)$~\cite{lovato:2013}. The Coulomb sum rule defined in \cref{eq:SL_def} only includes the inelastic contribution to $R_\alpha(\bm q,\omega)$, i.e., the elastic contribution represented by the second term on the right-hand side of \cref{eq:SL_gs}, where $|0;{\bm q}\rangle$ denotes the ground state of the nucleus recoiling with total momentum ${\bm q}$, has been removed.  This term is proportional to the longitudinal elastic form factor, which is given by
\begin{align}
	F_L(\bm q)=\frac{1}{Z} G_E^p(Q^2_{\rm el})\langle 0;{\bm q}|\mathcal O_L({\bm q}) |0 \rangle,
	\label{eq:elff}
\end{align}
where $Q^2_{\rm el}=\bm q^2-\omega_{\rm el}^2$, and $\omega_{\rm el}$ is the energy transfer corresponding to elastic scattering,
$\omega_{\rm el}=\sqrt{q^2+m_A^2}-m_A$ ($m_A$ is the mass of the target nucleus).

Neglecting the small spin-orbit contribution of \cref{eq:jL_def}, the Coulomb sum rule and the elastic form factor can be expressed as
\begin{align}
\label{eq:SL_cvmc}
    S_L(\bm q)&=\frac{1}{Z} \frac{1}{G_E^{p\,2}(Q_{\rm qe}^2)}\frac{1}{1+Q_{\rm qe}^2/(4 m^2)} \nonumber \\
    & \Big\{ G_E^{p\,2}(Q_{\rm qe}^2)\,\left[\tilde{\rho}_{pp}(q)+Z\right] \nonumber \\
    &        +G_E^{n\,2}(Q_{\rm qe}^2)\,\left[\tilde{\rho}_{nn}(q)+(A-Z)\right] \nonumber \\
    &        +2\,G_E^p(Q_{\rm qe}^2)\,G_E^n(Q_{\rm qe}^2)\,\tilde{\rho}_{np}(q) \nonumber \\ 
    & -\left[G_E^p(Q_{\rm qe}^2)\,\tilde{\rho}_p(q)+G_E^n(Q_{\rm qe}^2)\,\tilde{\rho}_n(q)\right]^2 \Big\},  \\[0.2cm]
	F_L(\bm q)&=\frac{1}{Z}\frac{G_E^p(Q_{\rm el}^2)\,\tilde{\rho}_p(q)+G_E^n(Q_{\rm el}^2)\,\tilde{\rho}_n(q)}{\sqrt{1+Q_{\rm el}^2/(4 m^2)}} \, ,
\end{align}
where $\tilde{\rho}_{N}(q)$ and $\tilde{\rho}_{\rm{NN}}(q)$ are the Fourier transform of the densities defined in \cref{eq:rho_N,eq:rho_NN}).

\begin{figure}[t]
	\centering
	\includegraphics[width=\linewidth]{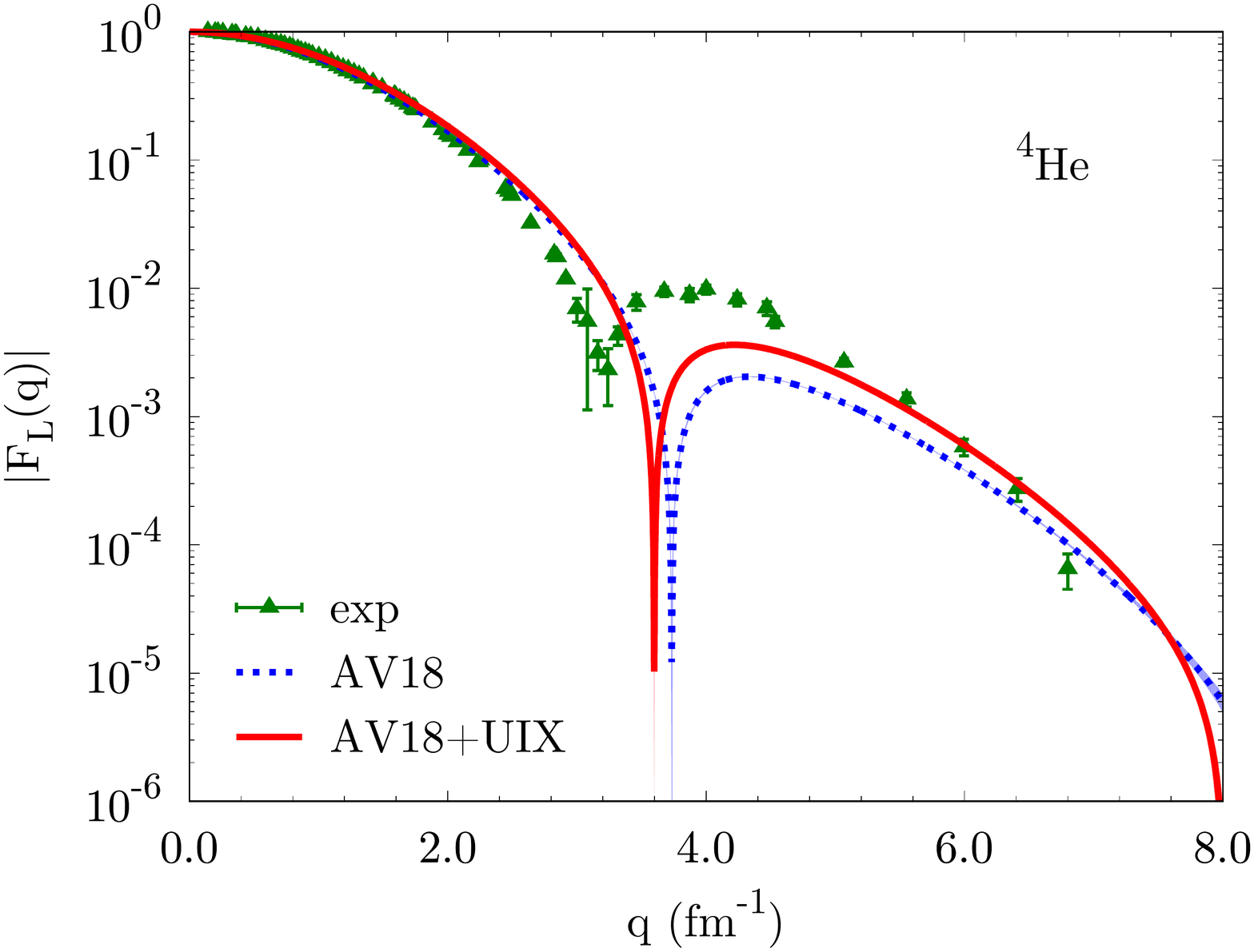}
	\caption[]{Longitudinal elastic form factors for $^4$He. Shaded areas indicate propagated Monte Carlo statistical errors in the Fourier transforms. Experimental data are from an unpublished compilation by I. Sick, based on Refs.~\cite{frosch:1967,erich:1968,mccarthy:1977,arnold:1978,ottermann:1985}.}
	\label{fig:ff_he4}
\end{figure}

\begin{figure}[b]
	\centering
	\includegraphics[width=\linewidth]{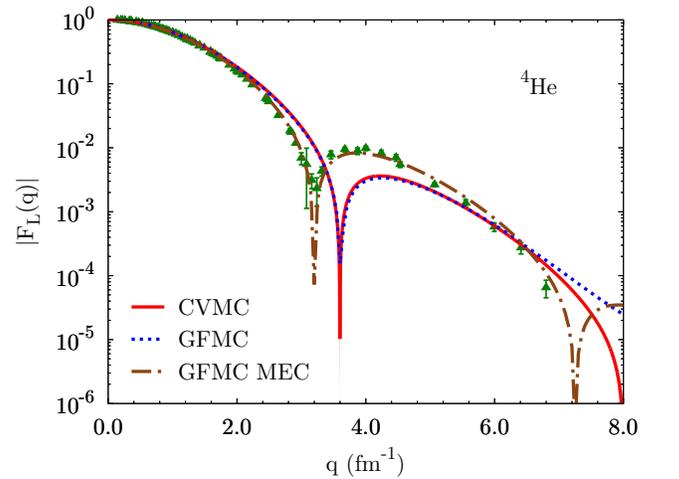}
	\caption[]{Longitudinal elastic form factors for $^4$He. Results employing the AV18+UIX potential are reported for CVMC and GFMC, the latter with and without MECs. Experimental results are the same as in Fig.~\ref{fig:ff_he4}.}
	\label{fig:ff_he4_comp}
\end{figure}

Here we compute the Coulomb sum rules and the elastic form factors of $^4$He, $^{16}$O, and $^{40}$Ca, to provide a useful benchmark for current and future analysis of electron-nucleus scattering data.  In particular, our results for $^{16}$O and $^{40}$Ca, when compared to experiment, should further elucidate the role of in-medium modification of the nucleon form factors.

The $^4$He longitudinal elastic form factor is compared to experimental data in \cref{fig:ff_he4}. Our theoretical results for the AV18 interaction significantly overpredict the diffraction minimum and maximum positions. Inclusion of the $3N$ force brings theory closer to experiment, but it is known that MECs are needed to further shift the peaks of the longitudinal elastic form factor to lower values of the momentum transfer and achieve agreement with experiment~\cite{marcucci:2016,carlson:2015}. This is shown in \cref{fig:ff_he4_comp} where the GFMC longitudinal elastic form factor with and without MEC contributions is displayed. Note that up to $\simeq 6\,\rm fm^{-1}$ the CVMC form factor perfectly matches the GFMC result obtained without MECs.

\begin{figure}[t]
	\centering
	\includegraphics[width=\linewidth]{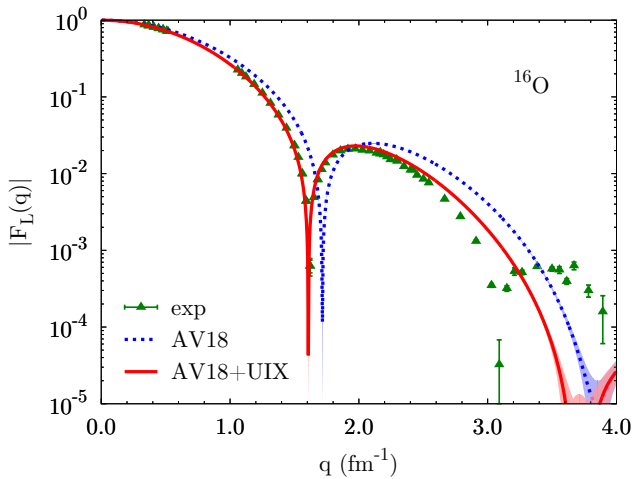}
	\caption[]{Longitudinal elastic form factors for $^{16}$O. Shaded areas indicate propagated Monte Carlo statistical errors in the Fourier transforms. Experimental data are from an unpublished compilation by I. Sick, based on Refs.~\cite{sick:1970,schuetz:1975,sick:1975}.}
	\label{fig:ff_o16}
\end{figure}

The longitudinal form factor of $^{16}$O is shown in \cref{fig:ff_o16}. The experimental data are well reproduced by our calculations once the $3N$ force is included. In analogy to $^{12}$C~\cite{lovato:2013}, it is plausible that two-body current contributions are negligible at low $q$, and become appreciable only for $q>3\,\rm fm^{-1}$. In fact, in the high-momentum region MECs interfere destructively with the one-body contributions, bringing theoretical prediction of $^{12}$C into closer agreement with experiment. This is consistent with the findings of Ref.~\cite{mihaila:2000}, where MECs improve the description of $^{16}$O experimental data above $2.5\,\rm fm^{-1}$.

\begin{figure}[b]
	\centering
	\includegraphics[width=\linewidth]{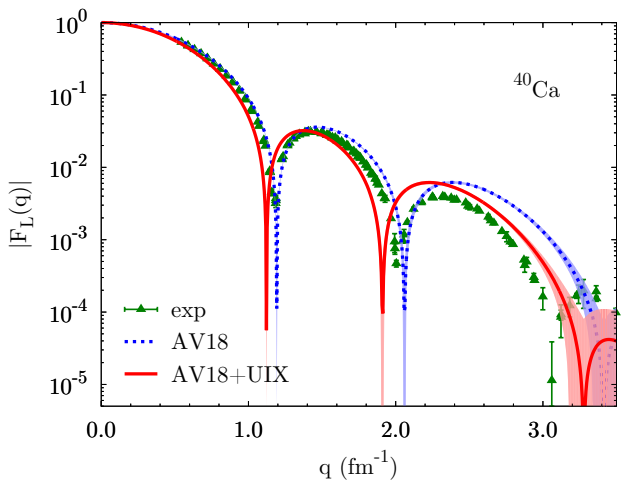}
	\caption[]{Longitudinal elastic form factors for $^{40}$Ca. Shaded areas indicate propagated Monte Carlo statistical errors in the Fourier transforms. Experimental data are from an unpublished compilation by I. Sick, based on Refs.~\cite{sinha:1973,sick:1975,sick:1979}.}
	\label{fig:ff_ca40}
\end{figure}

As for the $^{40}$Ca nucleus, a better agreement with experiments is achieved when AV18 only is present in the Hamiltonian (see \cref{fig:ff_ca40}). Assuming that, as for $^{12}$C and $^{16}$O, two-body current contributions have little effect for $q\leq 3\,\rm fm^{-1}$, we can infer that the UIX potential moves the diffraction peaks to excessively low values of $q$. This failure of the UIX interaction is directly related to the behavior of the point proton density displayed in \cref{fig:rhor_ca40}, where nucleons are pushed too far away from the center of mass when UIX is employed. 

\begin{figure}[t]
	\centering
	\includegraphics[width=\linewidth]{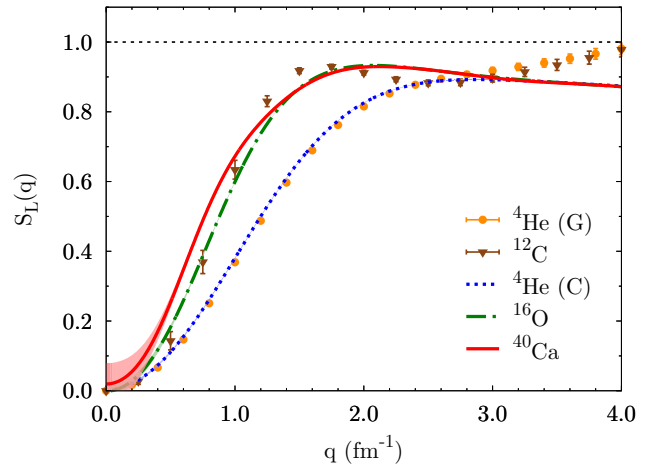}
	\caption[]{Coulomb sum rules for $A\leq40$. Symbols with statistical error bars show GFMC calculations employing the AV18+IL7 potential~\cite{lovato:2013}. The curves show CVMC results for AV18+UIX. Shaded areas indicate propagated Monte Carlo statistical errors in the Fourier transforms.}
	\label{fig:sl_av18+uix}
\end{figure}

The longitudinal sum rules of $^4$He, $^{16}$O, and $^{40}$Ca for AV18+UIX obtained from \cref{eq:SL_cvmc} are displayed in \cref{fig:sl_av18+uix}. The best GFMC estimates for $S_L(\bm q)$ in $^4$He and $^{12}$C~\cite{lovato:2013} are also shown for comparison (solid symbols). GFMC calculations have been carried out employing the AV18+IL7 potential and considering the full contribution of one- and two-body electromagnetic currents. The latter have only a relatively small effect on the longitudinal sum rule, mainly affecting the magnitude of the peak for $^{12}$C and the region above $3\,\rm fm^{-1}$. In this region, in addition to MECs, the discrepancies between CVMC and GFMC are due to the spin-orbit contribution in the charge operator, neglected in CVMC calculations but included in Ref.~\cite{lovato:2013}. In the large $q$ limit, the CVMC sum rules differ from unity because of relativistic corrections in the charge current, which gives the factor $1/(1+Q_{\rm qe}^2/(4m^2))$ of \cref{eq:SL_cvmc}. 

Extracting the Coulomb sum rules from the experimental response functions involves nontrivial difficulties. The experimental determination of $S_L(\bm q)$ requires measuring the associated $R_L(\bm q,\omega)$ from the inelastic threshold to infinity. However, inclusive electron scattering experiments can only explore the space-like region of the four-momentum transfer $\omega<q$. Therefore, a meaningful comparison between theory and experiment requires estimating the strength outside the region covered by electron-scattering experiments. Furthermore, the authors of Ref.~\cite{lovato:2016} have shown that the transitions to the low-lying states of $^{12}$C give significant contributions to $\langle 0|\mathcal O^\dagger_L(\bm q)\mathcal O_L(\bm q)|0 \rangle$ that are not present in the longitudinal response functions extracted from inclusive $(e,e^\prime)$ cross sections. Therefore, before comparing experiment with the present theory, which computes the sum rule of the total inelastic response rather than just the quasielastic one, these contributions have to be explicitly removed from the theoretical sum rule. In the $^{12}$C case, the transition form factors to $J^\pi=2^+$, $0_2^+$ (Hoyle), and $4^+$ states were taken from experiments. However, this approach is not suitable to the present work because of the large numbers of low-lying transitions of $^{16}$O and $^{40}$Ca. For this reason we refrain from reporting experimental data in \cref{fig:sl_av18+uix}.

\section{Conclusions}
\label{sec:conclusions}
A variational Monte Carlo analysis of the properties of three closed-shell nuclei, $^4$He, $^{16}$O, and $^{40}$Ca, has been performed. We employed the accurate phenomenological nuclear Hamiltonian AV18+UIX, which is capable of simultaneously describing two-nucleon bound and scattering states, the binding energy of $^4$He, and the saturation density of isospin-symmetric nuclear matter. The CVMC algorithm has been improved by including five-body terms in the cluster expansion of all the spin-isospin dependent correlations. Therefore, this work represents significant progress with respect to Ref.~\cite{pieper:1990,pieper:1992}, in which the older AV14+UVII Hamiltonian was employed, the cluster expansion was limited to four-body terms only, spin-orbit correlations were treated only at two-body cluster level, and other approximations were made in the construction of the wave function and in estimating the variational expectation values. 

In order to perform extensive searches for the optimal variational parameters in the multidimensional parameter space defined by the employed wave functions, we implemented in the CVMC program the open-source library for nonlinear optimization NLopt~\cite{johnson}. The accuracy of the optimized wave function has been tested against standard VMC and GFMC calculations for $^4$He using both AV18 and AV18+UIX, and against AFDMC results for $^4$He and $^{16}$O employing the AV6$^\prime$ potential.

We present results for the binding energy, charge radius, one- and two-body densities, single-nucleon momentum distribution, charge form factor, and Coulomb sum rule, fully accounting for the high-momentum components of the nuclear interaction. We find that the UIX three-body potential, known to be attractive for $A\leq 12$, becomes repulsive for $A\geq 16$. At variance with the $^4$He case, the addition of the UIX potential makes $^{16}$O and $^{40}$Ca less bound. This repulsive effect is not limited to the binding energies. In $^{16}$O and $^{40}$Ca nucleons are pushed far away from the center of mass when the $3N$ force is included, resulting in larger radii, broader densities, and a shift of the charge form factor diffraction peaks towards smaller momenta. Although relying on different interaction schemes, a similar behavior of three-body interactions is found in CC and IM-SRG calculations for medium-heavy nuclei (see~\cite{hagen:2014,hebeler:2015,hergert:2016} and references therein). We note that within CVMC there is no need to soften the $N\!N$ potential and to employ either the normal ordering procedure or a two-body density dependent approximation for the three-body force. 

Although the UIX three-nucleon interaction manifests a change in behavior---from attractive to repulsive---for $A\geq16$, it appears to provide a better description of radii, densities and charge form factors, of nuclei at least up to $A=16$. For instance, the charge radius and the position of the first peak in the longitudinal elastic form factor of $^{16}$O are better reproduced by the full AV18+UIX interaction than by the  AV18 potential alone. This is no longer true in $^{40}$Ca, where the inclusion of the $3N$ potential yields a too large charge radius and shifts the diffraction peaks of the charge form factor towards too small momenta. The experimental data for $^{40}$Ca lie in between the CVMC theoretical predictions for AV18 and AV18+UIX. The fact that the AV18+UIX Hamiltonian is not adequate to describe medium-mass nuclei is consistent with the deficiencies in the theoretical prediction of isospin-symmetric nuclear matter employing the same interaction. Although the correct saturation density is obtained, the binding energy per nucleon is too small~\cite{akmal:1998}. In this regard, as a follow up of this work, we will consider local $N\!N$ potentials recently derived in coordinate space within chiral perturbation theory~\cite{gezerlis:2013,gezerlis:2014,piarulli:2015,piarulli:2016,tews:2016,lynn:2016,logoteta:2016}. The latter are characterized by a spin-isospin structure analogous to the one of AV18+UIX so the CVMC can be straightforwardly extended to this class of interactions. It will be interesting to see whether local chiral effective field theory Hamiltonians provide a satisfactory description of $^{16}$O, $^{40}$Ca, and light nuclei.

\begin{figure}[b]
	\centering
	\includegraphics[width=\linewidth]{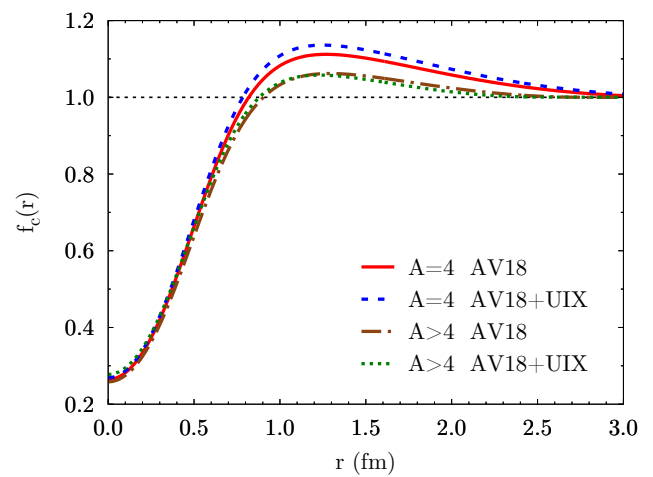}
	\caption[]{Central correlation functions for AV18 and AV18+UIX for $A=4$ and $A>4$ (the same two-body correlations have been employed in $^{16}$O and $^{40}$Ca; see text for details).}
	\label{fig:fc}
\end{figure}

\begin{figure}[t]
	\centering
	\includegraphics[width=\linewidth]{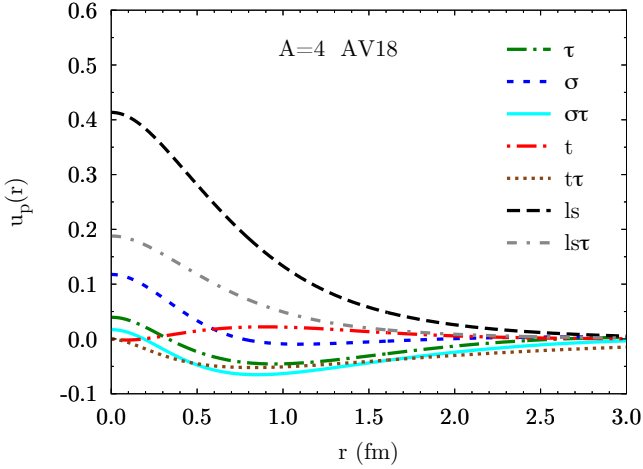}
	\caption[]{Radial correlation functions for $A=4$ and AV18.  $\tau,\,\sigma,\,\sigma\tau,\,t,\,t\tau,\,ls,\,ls\tau$ correspond to operators $p=2,\ldots,8$ in Eqs.~(\ref{eq:v6}) and (\ref{eq:v14}).}
	\label{fig:us_4_av18}
\end{figure}

\begin{figure}[b]
	\centering
	\includegraphics[width=\linewidth]{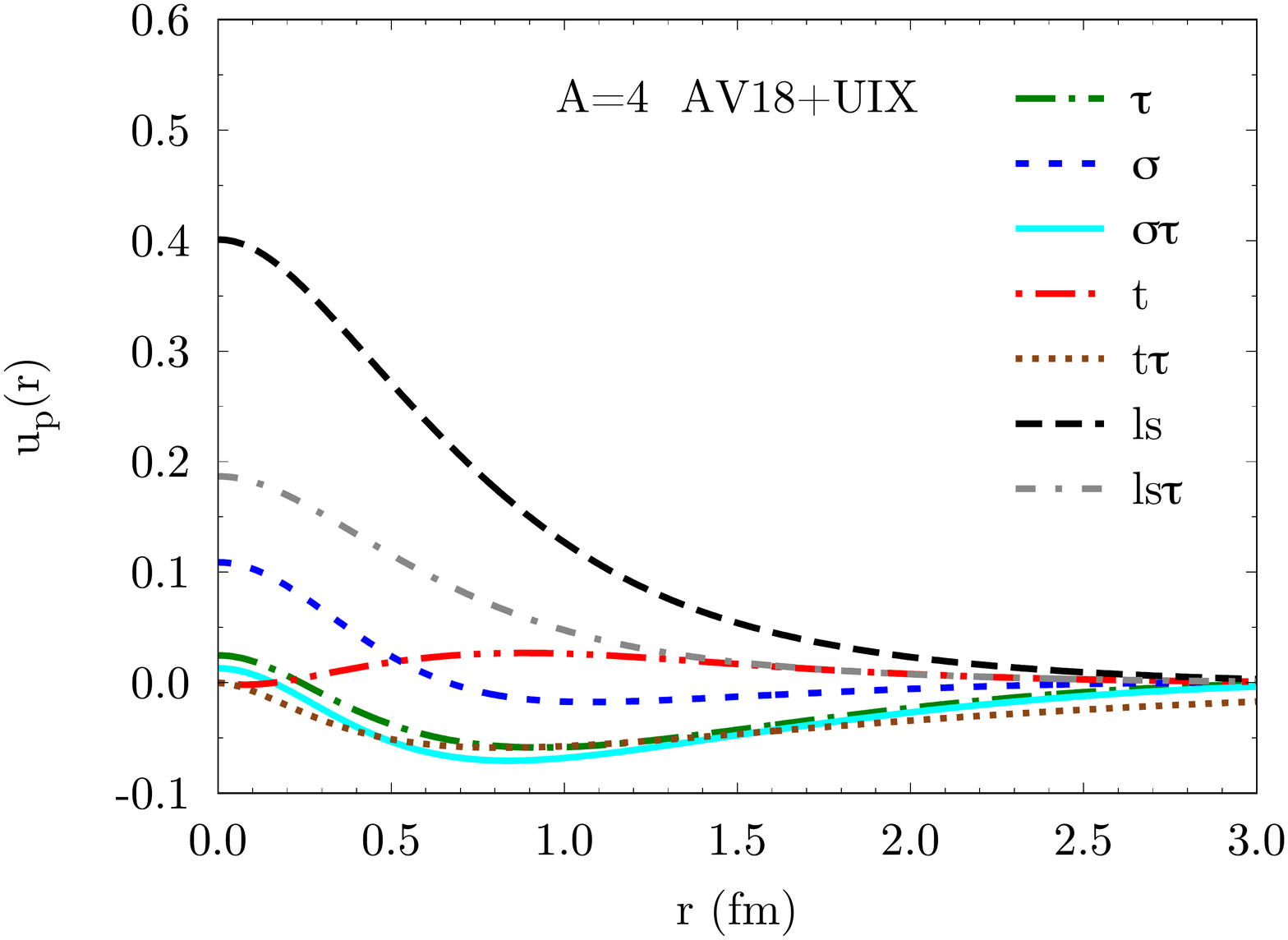}
	\caption[]{Radial correlation functions for $A=4$ and AV18+UIX.}
	\label{fig:us_4_av18+uix}
\end{figure}

We also computed the single-nucleon momentum distributions of $^{16}$O and $^{40}$Ca. These extend the VMC collection of Refs.~\cite{wiringa:2014,wiringa:rhok} obtained using realistic phenomenological Hamiltonians, which include both $N\!N$ and $3N$ interactions. Together with the inclusion of three-body and higher-order terms in the cluster expansion, this makes the calculations of $n(\bm k)$ accurate in both the high- and low-momentum regions. The universality of the tail of the momentum distribution, i.e., the independence of the high-momentum component upon the specific nucleus, has been confirmed for the selected interaction scheme. The momentum distributions are of immediate use for the studies of the high-momentum structure of nuclei, which includes the EMC effect and the analysis of short-range correlations in nuclei~\cite{arrington:2016}. For the latter, the analysis of two-nucleon momentum distributions derived employing realistic two- and three-body nuclear interactions will be of great interest. A future project will focus on the CVMC computation of two-nucleon momentum distributions in medium-heavy nuclei, extending the VMC collection of Refs.~\cite{wiringa:2014,wiringa:rhok12} and providing a comparison with the findings of Ref.~\cite{alvioli:2016}.

\begin{figure}[t]
	\centering
	\includegraphics[width=\linewidth]{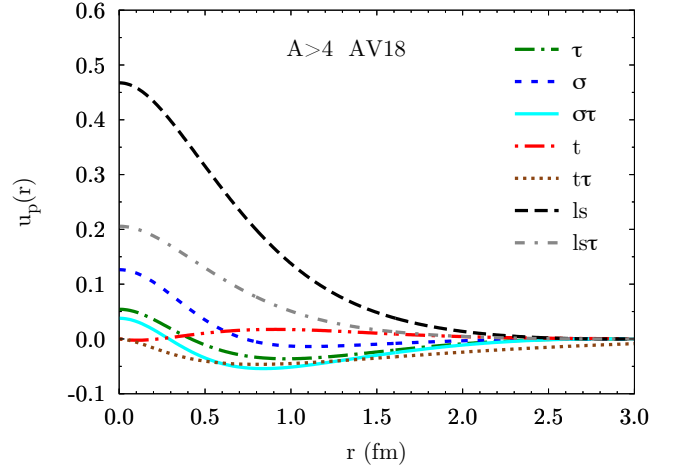}
	\caption[]{Radial correlation functions for $A>4$ and AV18.}
	\label{fig:us_16_av18}
\end{figure}

\begin{figure}[b]
	\centering
	\includegraphics[width=\linewidth]{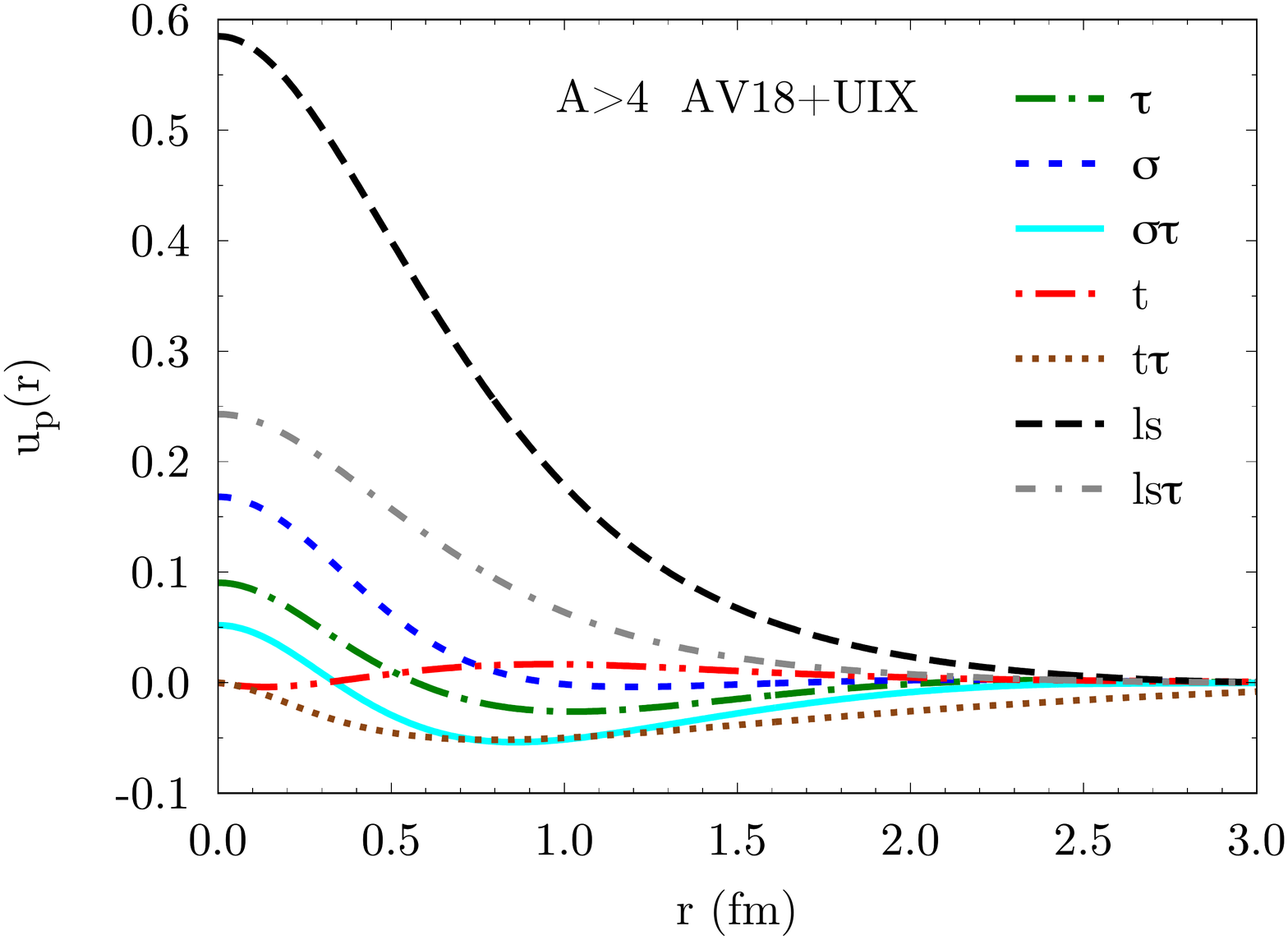}
	\caption[]{Radial correlation functions for $A>4$ and AV18+UIX.}
	\label{fig:us_16_av18+uix}
\end{figure}

We plan to employ the momentum distributions, and the average separation energies, computed in this work to evaluate the electroweak response functions of $^{16}$O and $^{40}$Ca in the impulse approximation, with particular emphasis on the role of three-nucleon forces, extending the study of Ref.~\cite{bacca:2009} to heavier nuclei. This will be relevant for neutrino-oscillation experiments, such as the Deep Underground Neutrino Experiment (DUNE)~\cite{DUNE}, and to elucidate quark and gluon effects in nuclei, which have long been actively sought, but never unambiguously identified.

\begin{table*}[t]
\centering
\caption{Variational parameters for $^4$He, $^{16}$O, and $^{40}$Ca.}
\label{tab:var}
\begin{tabular}{l c c c c c c}
\hline\hline
\multirow{2}{*}{param} & \multicolumn{2}{c}{$^4$He} & \multicolumn{2}{c}{$^{16}$O} & \multicolumn{2}{c}{$^{40}$Ca} \\
& AV18 & AV18+UIX & AV18 & AV18+UIX & AV18 & AV18+UIX \\
\hline
$V_S$                  & $-47.824\,\rm MeV$   & $-51.459\,\rm MeV$    & $-44.860\,\rm MeV$   & $-43.032\,\rm MeV$   & $-47.261\,\rm MeV$   & $-45.606\,\rm MeV$ \\ 
$R_s$                  & $2.174\,\rm fm$      & $2.039\,\rm fm$       & $3.325\,\rm fm$      & $3.542\,\rm fm$      & $4.592\,\rm fm$      & $4.872\,\rm fm$ \\
$a_s$                  & $0.371\,\rm fm$      & $0.340\,\rm fm$       & $0.439\,\rm fm$      & $0.641\,\rm fm$      & $0.690\,\rm fm$      & $0.930\,\rm fm$ \\
$\alpha_s$             & $0.126$              & $0.285$               & $-0.056$             & $-0.091$             & $-0.056$             & $-0.091$ \\ 
$\rho_s$               & $1.643\,\rm fm$      & $1.131\,\rm fm$       & $1.847\,\rm fm$      & $1.009\,\rm fm$      & $1.847\,\rm fm$      & $1.009\,\rm fm$ \\ [0.2cm]
$k_F$                  & $1.864\,\rm fm^{-1}$ & $1.7036\,\rm fm^{-1}$ & $1.604\,\rm fm^{-1}$ & $1.309\,\rm fm^{-1}$ & $1.604\,\rm fm^{-1}$ & $1.309\,\rm fm^{-1}$ \\
$\alpha$               & $0.759$              & $0.745$               & $0.787$              & $0.893$              & $0.787$              & $0.893$ \\
$\beta_c$              & $0.976$              & $0.994$               & $1.078$              & $1.172$              & $1.078$              & $1.172$ \\
$\beta_t$              & $1.206$              & $1.341$               & $1.130$              & $1.194$              & $1.130$              & $1.194$ \\ 
$d_{\rm S}$            & $3.361\,\rm fm$      & $3.539\,\rm fm$       & $2.787\,\rm fm$      & $2.502\,\rm fm$      & $2.787\,\rm fm$      & $2.502\,\rm fm$ \\
$d_{\rm P}$            & $4.711\,\rm fm$      & $4.039\,\rm fm$       & $2.867\,\rm fm$      & $3.212\,\rm fm$      & $2.867\,\rm fm$      & $3.212\,\rm fm$ \\
$d_t$                  & $6.449\,\rm fm$      & $6.716\,\rm fm$       & $4.655\,\rm fm$      & $4.312\,\rm fm$      & $4.655\,\rm fm$      & $4.312\,\rm fm$ \\ [0.2cm]
$t_1$                  & $5.792$              & $5.769$               & $5.165$              & $4.097$              & $5.165$              & $4.097$ \\
$t_2$                  & $4$                  & $4$                   & $4$                  & $4$                  & $4$                  & $4$ \\
$t_3$                  & $0.127\,\rm fm^{-1}$ & $0.117\,\rm fm^{-1}$  & $0.252\,\rm fm^{-1}$ & $0.202\,\rm fm^{-1}$ & $0.252\,\rm fm^{-1}$ & $0.202\,\rm fm^{-1}$ \\ [0.2cm]
$\varepsilon_{2\pi,A}$ &                      & $-9.60\cdot10^{-4}$   &                      & $-8.77\cdot10^{-4}$  &                      & $-8.77\cdot10^{-4}$ \\
$\varepsilon_R$        &                      & $-8.22\cdot10^{-4}$   &                      & $-7.87\cdot10^{-4}$  &                      & $-7.87\cdot10^{-4}$ \\
$\eta$                 &                      & $0.693$               &                      & $1.005$              &                      & $1.005$ \\
$c_y$                  &                      & $1.337\,\rm fm^{-2}$  &                      & $1.601\,\rm fm^{-2}$ &                      & $1.619\,\rm fm^{-2}$ \\
$c_t$                  &                      & $1.811\,\rm fm^{-2}$  &                      & $1.616\,\rm fm^{-2}$ &                      & $1.734\,\rm fm^{-2}$ \\
\hline\hline
\end{tabular}
\end{table*}

We computed the Coulomb sum rules for closed-shell nuclei ranging from $A=4$ to $A=40$. Our calculations show very little $A$ dependence of the sum rules for $A \geq 12$ for momentum transfers as low as $1\,\rm fm^{-1}$. These results are also consistent with the recent GFMC calculation for $^{12}$C~\cite{lovato:2016}.  

Another future project will be to examine closed-shell nuclei $+/-$ one nucleon, e.g., $^{15}$N, $^{15}$O, $^{17}$O, $^{17}$F, to study various properties such as spin-orbit splitting, which was previously evaluated in $^{15}$N using CVMC in Ref.~\cite{pieper:1993}, charge-symmetry breaking~\cite{wiringa:2013}, and $\beta$ decay.

\acknowledgments{We thank I. Sick for providing us with the compilation of the experimental longitudinal elastic form factors and for the interpretation of the data. We are also thankful to J. Carlson and S. Gandolfi for insightful discussions. This work was supported by the U.S. Department of Energy, Office of Science, Office of Nuclear Physics, under the FRIB Theory Alliance Grant Contract No. DE-SC0013617 titled ``FRIB Theory Center---A path for the science at FRIB'' (D.L.), under the NUCLEI SciDAC grant (D.L. and A.L.), and under Contract No. DE-AC02-06CH11357 (A.L., S.C.P, and R.B.W). Computing time was provided by Los Alamos Open Supercomputing via the Institutional Computing (IC) program, by the National Energy Research Scientific Computing Center (NERSC), which is supported by the U.S. Department of Energy, Office of Science, under Contract No. DE-AC02-05CH11231, and by the Laboratory Computing Resource Center (LCRC) at Argonne National Laboratory.}

\section*{Appendix: Wave function details}
\Cref{fig:fc,fig:us_4_av18,fig:us_4_av18+uix,fig:us_16_av18,fig:us_16_av18+uix} and \cref{tab:var} provide the radial correlation functions and all the variational parameters for the systems under study for both AV18 and AV18+UIX.

%

\end{document}